\documentclass[preprint2]{emulateapj}  
\usepackage{graphicx}

\shorttitle{CME rate during solar cycle 23}
\shortauthors{E. Robbrecht et al.}

\begin{document}

\title{Automated LASCO CME catalog for solar cycle 23:\\ are CMEs scale invariant?}

\author{E. Robbrecht\altaffilmark{1,2,3} , D. Berghmans\altaffilmark{1} and R.A.M. Van der Linden\altaffilmark{1}}
\altaffiltext{1}{SIDC - Royal Observatory of Belgium, Ringlaan 3, 1180 Brussel, Belgium}
\altaffiltext{2}{Naval Research Laboratory, 4555 Overlook Ave SW, Washington DC 20375, USA}
\altaffiltext{3}{CEOSR, George Mason University, VA 22030, USA} 

\email{Eva.Robbrecht.ctr.be@nrl.navy.mil}


\begin{abstract}
In this paper we present the first automatically constructed LASCO CME catalog, a result of the application of the Computer Aided CME Tracking software (CACTus) on the LASCO archive during the interval September 1997 -  January 2007. We have studied the CME characteristics and have compared them with similar results obtained by manual detection (CDAW CME catalog).  On average CACTus detects less than 2 events per day during solar minimum up to 8 events during maximum, nearly half of them being narrow ($< 20\degr$). Assuming a correction factor, we find that the CACTus CME rate is surprisingly consistent with CME rates found during the past 30 years. The CACTus statistics show that small scale outflow is ubiquitously observed in the outer corona. The majority of CACTus-{\it only} events are narrow transients related to previous CME activity or to intensity variations in the slow solar wind, reflecting its turbulent nature. A significant fraction (about 15\%) of CACTus-{\it only} events were identified as independent events, thus not related to other CME activity. The CACTus CME width distribution is essentially scale invariant in angular span over a range of scales from 20 to $120\degr$ while previous catalogues present a broad maximum around $30\degr$. The possibility that the size of coronal mass outflows follow a power law distribution could indicate that no typical CME size exists, i.e. that the narrow transients are not different from the larger well-defined CMEs. 

\end{abstract}


\section{Introduction}

In this paper we discuss an attempt to quantify the detection of coronal mass ejections (CMEs). Coronal mass ejections (CMEs) are episodic expulsions of mass and
magnetic field from the solar corona into the interplanetary medium. A classical CME carries away some $10^{15}$~g of coronal mass and can liberate energies of $10^{23}-10^{25}$ J \citep{1985JGR....90.8173H,2002svco.conf...91V}. In broad band white-light coronagraphic images CMEs are seen as bright features moving radially outward. Building a CME catalog basically means listing all occurrences of events, defined as CMEs. CMEs can be very bright and often show evidence of magnetic structure (e.g. twisted flux-rope), but  sometimes, no discernible structure is present or the intensity enhancement is only very weak compared to the background corona (e.g. due to projection effects), which makes it very hard to detect and characterize them. 
Application of the automated CME detection software on the LASCO archive (see next section for a description of the Software) shows a picture of coronal activity that corresponds well to the variety of CME-types presented in \cite{1985JGR....90.8173H}. 

\begin{figure}[h]
\centering
\includegraphics[width=\columnwidth]{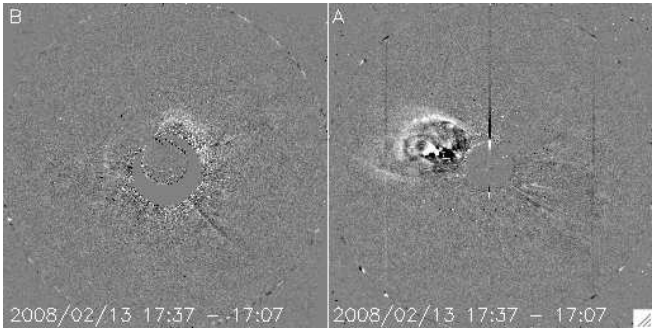}
\caption{STEREO/COR2 image pair in running difference (left: B, right: A, separation angle: $45\degr$) illustrating the influence of projection effects on the appearance of a CME. The CME was observed on Feb 13 2008 as a near-limb CME by the A spacecraft and a faint partial halo in the B spacecraft. \label{fig:stereo-pair}}
\end{figure}

After 3 decades of coronagraphic observations, the statistical properties of CMEs are relatively well known. CME angular span, speeds, latitudes and masses have been measured and statistically analyzed \citep[e.g.][and references therein]{2004JGRA..10907105Y, 2007AdSpR..40.1042C}. In contrast to this huge amount of observations and studies of CMEs, there remain a number of unresolved issues and their physics is not well understood, especially their initiation mechanism. Ever since the start of CME observation, several events had an `unclear' status and up to date a large fraction of the observations does not fit in the `flux rope CME' picture. Do they appear differently because of effects of projection and Thomson scattering? \cite{2004A&A...422..307C} have shown how big the impact of projection effects can be. Currently, STEREO/SECCHI \citep{2008SSRv..136...67H} observations show unambiguous evidence that coronagraphic observations only show a 2D reflection of the whole 3D corona. For example a bright well defined CME was observed in the A spacecraft on Feb 13 2008. With the B spacecraft, $45\degr$ separated from A, only a faint partial halo was detected (see Fig. \ref{fig:stereo-pair}). Undoubtedly, STEREO will advance greatly our insight in these effects. Also instrumental sensitivity influences what we see. The `double spike' events, as classified by \cite{1985JGR....90.8173H}, were believed to be part of one event (and hence listed together) consisting of 2 legs connected by a faint arch. The arch was too faint to be observed by Solwind, but was observed by SMM \citep{1980SoPh...65...91M}. So, what we call `background outflow' might actually be an erupting magnetic structure, containing e.g. a mini-flux rope (mini referring to angular sizes smaller than $20\degr$). Could the `single spikes' simply be double spikes of which only one leg is visible? 

Small scale variations are more numerous than large `structured' events. Are they the signatures of magnetic instabilities seen as episodic expulsions of mass? High resolution STEREO/EUVI images show small dimmings across the solar disk, covering the quiet sun. Undoubtedly, part of this activity is seen higher in the corona. Just as the quiet sun is not really quiet, the slow solar wind is not merely a quiet steady flow, but a flow with turbulent nature. Where does the turbulence end and the `foreground' activity start? At solar active times, a wealth of outward moving brightness features is observed mostly as narrow transients, complementary to the well-distinct CMEs. Is it possible to draw an imaginary line between `real CMEs' and `small discrete outflow' on physical grounds, or does there exist a continuum from large bright CMEs to small unimportant events?  E.g. are jets along streamers simply the larger ``blobs" observed by \cite{1997ApJ...484..472S} or are they at the lower range of CMEs? The problem of  the inclusion of `narrow' events in catalogs is not new and dates from pre-LASCO observations. In an examination of Solwind coronagraphic images, \cite{1985JGR....90.8173H} had ``no trouble agreeing that large bright CMEs were significant events. The question became whether to include all faint or very narrow CMEs in our analysis."  

The first study to provide a statistical view of the properties of CMEs observed by LASCO during 1996-1998 is given by \cite{2000JGR...10518169S}. In this study it is explicitly mentioned that ``(1) the polar microjets reported by \cite{1997SoPh..175..571M} and  (2) the small inhomogeneities that may trace out the low latitude acceleration of the slow solar wind \citep{1997ApJ...484..472S} are both excluded." The authors confirm that these marginal events satisfy the observable definition of coronal mass ejections, but they are excluded from the statistical study.  

\begin{figure}\centering
\includegraphics[width=\linewidth]{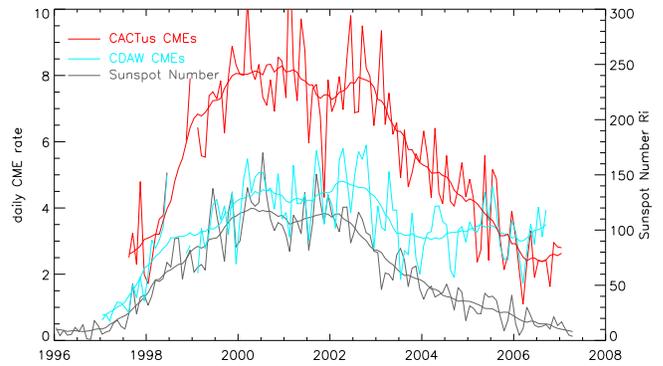}
\caption{The daily SOHO LASCO CME rates for cycle 23 ({\it thin curves:} smoothed per month, {\it thick curves:} smoothed over 13 months) from 1997 trough 2006, extracted from the CACTus (red) and the CDAW (blue) CME catalog. As a reference we have overplotted the monthly and monthly smoothed sunspot number (grey) produced by the SIDC - Royal Observatory of Belgium. The CME rates have been corrected for duty cycle (see text for details).
\label{fig:cmerate}}
\end{figure}


\begin{deluxetable*}{lcccccccccc}
\tablecaption{Average daily CME rates derived from LASCO (cycle 23) duty cycle corrected.
\label{table:cmerate}}
\tablewidth{0pt}
\tablehead{Data & \colhead{1997}  &  \colhead{1998}  &  \colhead{1999}  &  \colhead{2000}  &  \colhead{2001}  &  \colhead{2002}  &  \colhead{2003}  &  \colhead{2004}  &  \colhead{2005}  &  \colhead{2006}  } 
\tablecolumns{11}
\startdata
CACTus & 3.0 & 4.2 & 7.2 & 8.1 & 7.8 & 7.8 & 6.3 & 4.9 & 3.7 & 2.4 \\
CDAW    & 1.1 & 3.1 & 3.4 & 4.6 & 4.3 & 4.7 & 3.3 & 3.2 & 3.5 & 3.1 \enddata
\end{deluxetable*}

In the next section we describe the composition of the CACTus catalog and the available data. Thereafter, we focus on the CME rate during cycle 23 (section \ref{sec:cmerate}) and discuss the statistics of the CME parameters (section \ref{sec:CMEparam}). Particular attention is given to the small ejections and outflow in the discussion section (section \ref{sec:discuss}).

\section{Composition of the catalog}

Based on the Computer Aided CME Tracking software (CACTus), we have constructed an objective CME catalog based on LASCO data \citep{1995SoPh..162..357B} spanning the period September 1997 - January 2007. We refer to it as the `CACTus CME catalog' and it can be found online at http://sidc.be/cactus. The CACTus software package was first reported in \cite{2002svco.conf...85B} and is extensively described in \cite{2004A&A...425.1097R}. CACTus detects CMEs in height-time maps constructed from LASCO C2/C3 images. CMEs are seen as inclined lines in height-time maps and are detected using the Hough transform. The method has 2 inherent limitations: (1) only {\it radial} motion can be detected and (2) no acceleration can be measured since CACTus detects {\it straight} lines in the height-time maps. At present, CACTus measures the following parameters: first time of appearance in C2, CME width, principal direction (defined as the middle direction of the CME) and a linear speed profile along the angular span of the CME. To limit computation time and false detections, we have set 3 criteria for the selection of CMEs. Only detections with plane-of-sky-speeds between 100 and 2100 km/s (slow CMEs require most computation time since they need many images to travel through the C2/C3 FOV (field of view); the errors on the speed measurements become large for faster CMEs), with an integrated $\Delta I/I$ ridge-intensity (in the height-time space) above a fixed threshold and with an angular span $\ge 10\degr$ are retained. 

Prior to preprocessing, the images are tested for their reliability. This step is performed in order to limit the amount of false detections due to corrupt images. They arise e.g. from dust particles or small debris flying in front of the telescope just at the time an image was taken, from highly deviating exposure times and from errors in data acquisition, transmission and reconstruction.  During the first months of the mission, only the equatorial region for the FOV was transmitted. This style of image compression was gradually decreased and abandoned in September 1997. Moreover, the nominal cadence of both C2 and C3 was only 1 image per hour (compared to resp. 3 and 2 per hour). For these reasons the current data set used for our long-term analysis runs from September 1997 until January 2007. Nominal observations have been interrupted as a consequence of exceptional satellite problems\footnote[1]{ {\small http://sohowww.nascom.nasa.gov/about/Recovery/docs/ index.html}}. A 3 months data gap occurred in 1998 from 24 June to 22 October due to an unexpected loss of contact with the spacecraft. Subsequent failure of all three gyroscopes caused an interruption from 21 December 1998 to 6 February 1999. A third data gap occurred in June 2003, when SOHO's main antenna became stuck. This problem was overcome and nominal observations resumed on July 10. Additionally, regular gaps of a few days through the whole mission's lifetime occur during the SOHO `keyhole periods'. 

\section{CME rate during cycle 23}\label{sec:cmerate}
\placefigure{fig:cmerate}
\placetable{table:cmerate}

Figure \ref{fig:cmerate} shows the daily CACTus CME rate for cycle 23 in red, with the International Sunspot Number \citep[e.g.][]{2004SoPh..224..113V} superimposed in grey as solar cycle reference. We have also plotted the daily CDAW CME rates in blue \citep{2004JGRA..10907105Y}. It is available online and is widely used by the solar community as a reference LASCO CME catalog. The {\it average} daily values for CACTus and CDAW are given in Table \ref{table:cmerate}. The CME rates that we report in this paper have been corrected for instrument duty cycle. We applied a different correction to the CACTus CME rates and the CDAW rates, because CACTus does not accept all images. For CACTus we deduced the number of effective observation days from the actual images we used as input, by subtracting all data gaps that were larger than 12 hours. We did this based on the C2 data alone, C2 data gaps overlap greatly with C3 data gaps. For correcting the CDAW CME rate, we used a file containing the C2 door closing times and subtracted all closing times from the total month-time. For each catalog we then scaled the number of CMEs counted during that month to the calculated number of observation days. We have applied a smoothing function on the monthly CME and sunspot rates by computing a boxcar (running) average over a smoothing window of 13 months. Our findings are:

\begin{figure}\centering
\includegraphics[width=\columnwidth]{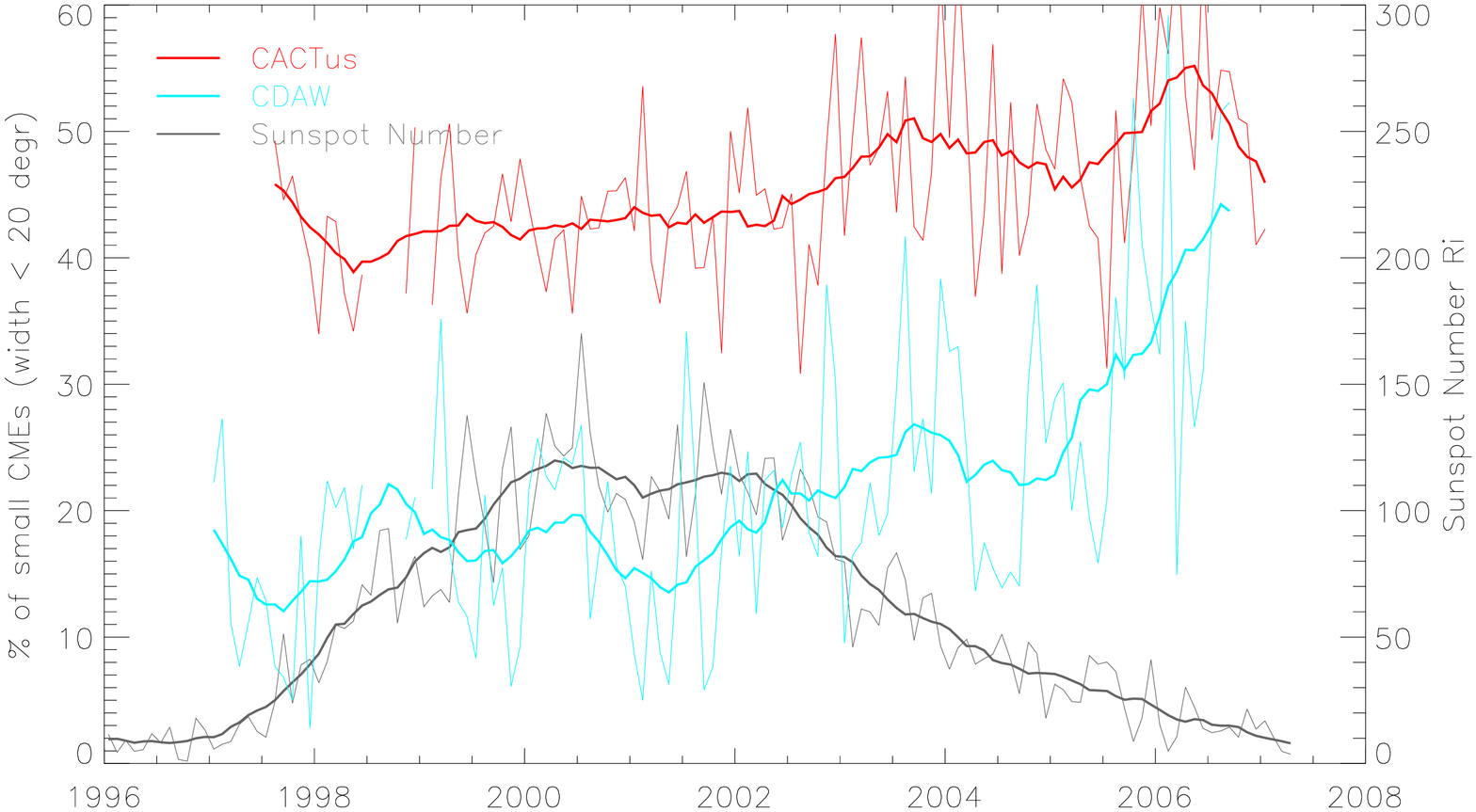}
\caption{Percentage of narrow CMEs (width smaller than $20\degr$) compared to the total number of CMEs as listed by the CACTus (red) and CDAW (blue) catalog. As a reference to the solar cycle we have plotted in grey the monthly and monthly smoothed sunspot number produced by the SIDC - Royal Observatory of Belgium.\label{fig:cmerate2}}
\end{figure}

{\bf [1] Solar cycle effects}\,\,The smoothed CACTus CME rate (fig. \ref{fig:cmerate}, thick red curve) confirms the pre-SOHO observation that the CME rate follows the solar cycle \citep{1994JGR....99.4201W}, here represented by the Sunspot Number. Also the Gnevychev gap \cite[GG][]{1967SoPh....1..107G}, the dip in the maximum phase of solar activity, is well retrieved in the CACTus curve.  Only the general trend is correlated, on short timescales the CME rate and sunspot number are not well correlated. The daily CACTus CME rate averaged per year increases roughly with a factor 4 from minimum to maximum. This factor is more or less stable for the different sizes of CMEs. On average there are $\sim 2$ CACTus events per day during solar minimum and $\sim 8$ events during solar maximum. Figure \ref{fig:cmerate2} shows that nearly half of the CACTus detections are narrow events ($< 20^{\circ}$).  
 
{\bf [2] CACTus rate is higher than CDAW rate}\,\, As can be seen in figure  \ref{fig:cmerate} the CACTus CME rate is much higher than the CDAW rate (blue curve) for most of cycle 23. CACTus detects all bright outward radial motion independent of morphology or the presence of other activity. An observer will generally not list outflow activity in the aftermath of a large bright CME. In the discussion section, we will focus on the detection and quantization of coronal activity in general. The large discrepancy between the two CME curves is most pronounced during solar maximum years, but it is also present during other years. The flat CDAW curve in the decaying phase of the cycle ($2004-2007$) is very surprising in Fig.\ref{fig:cmerate}. Since CACTus measures a systematic decrease from maximum to minimum and also the sunspot number decreases continuously we do not interpret the CDAW flat rate as solely due to physical effects. Instead, different criteria used by different personnel could be the cause of differently populating the CME catalog \citep[see][for a discussion]{2008SoPh..249..355K}. This however is unfortunate for CME statistics and shows the need for automated measurement of CME-activity in the corona, in which the introduced biases or consistent for the whole observation. 


\begin{figure*}[t]\centering
\includegraphics[width=.49\linewidth]{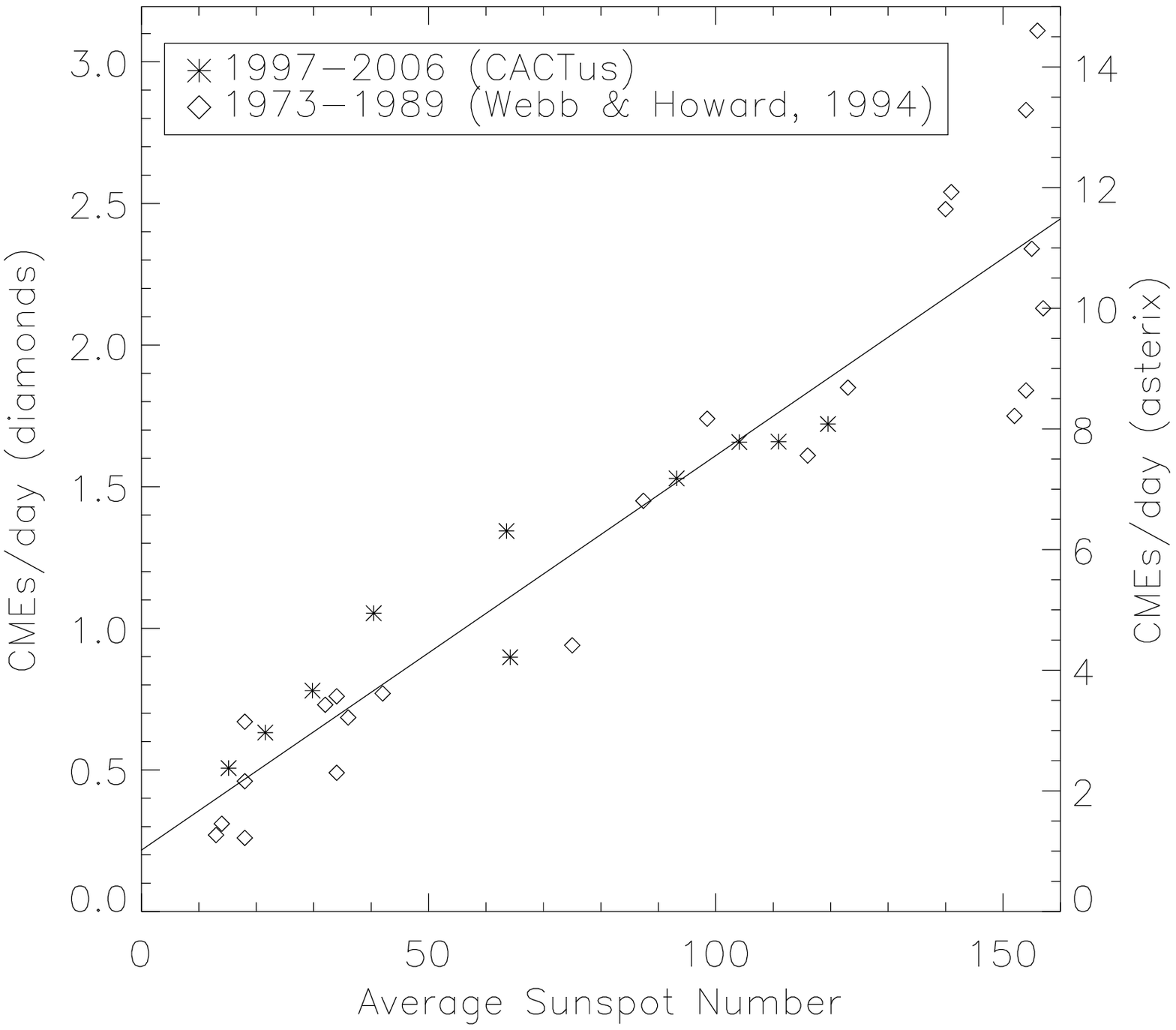}
\includegraphics[width=.49\linewidth]{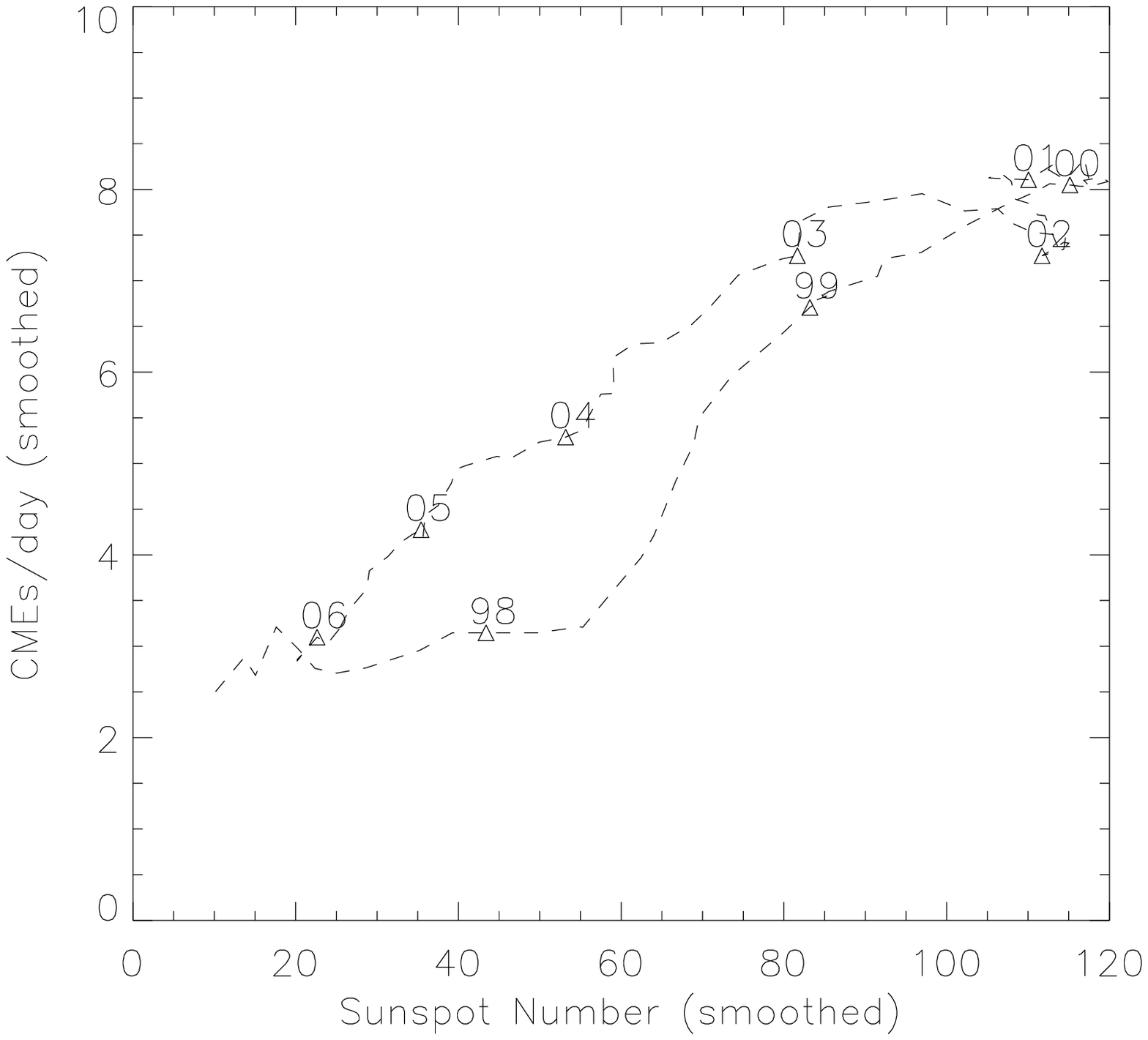}
\caption{{\it Left:} Daily CME rate vs sunspot number, both averaged per year. The asterixes refer to rates for the cycle 23 derived from CACTus (see Table \ref{table:cmerate}). Its absolute scale is shown on the right $y$-axis. The daily CME rates derived by \cite{1994JGR....99.4201W} are plotted with diamonds. Its absolute scale is shown on the left $y$-axis. A scaling factor of $\sim4.7$ applies between the CACTus and the Webb \& Howard rates. {\it Right:} Smoothed daily CACTus CME rate vs smoothed sunspot number. The triangles denote the start of every year.
\label{fig:cme_vs_ssn}}
\end{figure*}

{\bf [3] CME cycle lags sunspot cycle}\,\, CME activity of cycle 23 shows a significant peak delay with respect to the sunspot cycle (see Fig.~\ref{fig:cmerate}). Focussing on the monthly averaged curves, we find a lag-time varying from 6 months (max peaks) to 1 year (Gnevychev Gap). The CME rate during cycle 23 thus tracks the solar activity cycle in amplitude but phase-shifted. Since this effect was not clearly present in the activity rates of cycle 21-22  \citep{1994JGR....99.4201W}, this is possibly a peculiarity to cycle 23. The phenomenon of time-delay, has been observed in several other activity indicators.  Chromospheric and coronal emission lines show delays of 1 to 4 months w.r.t. the sunspot index \citep{1983JGR....88.9883D,1994SoPh..150..347B} and time-lags of 10 to 15 months are found for flare rates \citep{2001IAUS..203..125O,2003SoPh..215..111T}. The mechanism leading to these and similar delays is not understood. An obvious remark to make here is that sunspots only reflect part of the source regions of CMEs. Several studies found that the majority of CMEs for which on-disk signatures could be observed are related to filament/prominence eruptions \citep[e.g.][]{1979SoPh...61..201M,1987SoPh..108..383W}. Nevertheless, when treating the sunspot number as proxy for the (long-term) solar cycle (i.e. not as a count of individual source regions), the observed time-delays give an idea of the time needed to build up the necessary conditions for coronal activity.  


{\bf [4] CME rate is consistent with past cycles} \,\, Figure \ref{fig:cme_vs_ssn} ({\it left}) shows the daily CME rate vs sunspot number (SSN) both averaged per year. The asterixes refer to rates for the current cycle (cycle 23) derived from CACTus, its absolute scale is shown on the right $y$-axis. The daily CME rates derived by \cite{1994JGR....99.4201W} are plotted with diamonds, its absolute scale is shown on the left $y$-axis. The Webb \& Howard rates are corrected for duty cycle and instrumental visibility and are based on data from Skylab, Helios (zodiacal light photometer data), Solwind and SMM. In total, they cover the period  between 1973 - 1989. The absolute rates for cycle 23 are much higher than those reported for previous cycles. This is due to the better instrument sensitivity, the enormous dynamic range of LASCO, the much larger field of view and the more uniform coverage of data over a long period of time.  Additionally, the CACTus detection system has higher detection sensitivity than manual detection, i.e. it picks up all radial outflow that exceeds the thresholds set for brightness and angle. By applying a simple scaling factor of $\sim4.7$ to the previous CME rates, we could fit them to the CACTus scale, or vice versa.  Given the fact that these data points are derived from different instruments, using different techniques (manual vs automatic) over several solar cycles, these points match extremely well. Once again, this confirms that long-term CME activity is a function of the solar activity, here represented by the SSN. It can be seen that the current cycle was less strong than the previous cycles, the SSN only reaches $\sim 120$, whereas for the 2 previous cycles a maximum of $\sim 160$ was retrieved. Likewise, the ratio of CME-rates between solar maximum and minimum is $\sim 4$ for the current cycle, which is smaller than it was for the previous cycle where the ratio was on average larger than 5. From this comparison we estimate that the CME-activity was lower during cycle 23 compared to the previous cycle, despite the fact that the {\it absolute} CME rates were higher.

{\bf [5] CME rate rises faster than it decays} \,\,In Figure  \ref{fig:cme_vs_ssn} ({\it right}) we plot the smoothed CACTus daily CME rate vs the smoothed sunspot number. From this plot it can be inferred that just like the SSN the CME rate rises steep and decays slowly after solar maximum. This means that for the same number of sunspots more CMEs are produced during the decaying phase. This doesn't necessarily mean that these sunspots are more active, it could also mean that more CMEs erupt from non-sunspot regions. This will be discussed in a subsequent paper (Robbrecht et al. 2008, {\it in prep}).

\section{Statistics of CME parameters}\label{sec:CMEparam}

\begin{figure}[h]\centering
{{\includegraphics[width=\columnwidth]{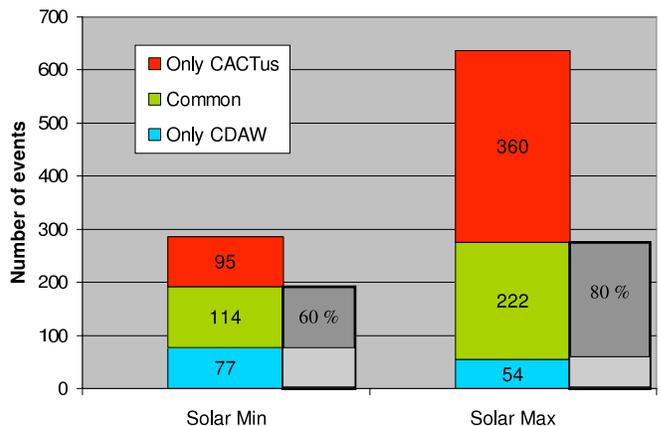}}}
\caption{Graph illustrating the CACTus-CDAW correspondence for two
selected samples. It is based on two months during solar minimum
(1998) and two months during solar maximum (2000). The grey boxes
only cover the CDAW CMEs. Only 60\% during solar minimum and 80 \%
during solar maximum of these particular events could be connected
to a CACTus CME detection (green). Beside these, CACTus has detected
many other events, which are not present in the CDAW catalog
(red).\label{fig:sample}}
\end{figure}

\placefigure{fig:sample}
In this section we discuss the statistics of the CACTus CME parameters and compare them with the CDAW statistics. We also try to estimate the effect of measurement method on the different CME parameters (starting time, principal angle, angular width, speed) by comparing the measurements for a sample of common events (i.e. present in both catalogs).  
This sample was chosen large enough (336 events), such that the results are statistically significant and expandable towards the whole catalog. CME occurrence depends on the solar cycle, therefore we have selected two different sub-samples, one representing solar minimum (1998 Feb and May) and the other solar maximum (2000 Apr and Aug). For each day in each month we have plotted the detections on an angle-time map and have visually inspected the LASCO movies in order to decide which entries are describing the same event. This lead to 114 common events for the minimum sample and 222 common events for the maximum. Figure \ref{fig:sample} gives an overview of the CACTus-CDAW correspondence for the two selected periods. During solar maximum, 80\% of the CDAW CMEs had a corresponding CACTus detection, but only 60\% during solar minimum. We attribute this lower value to the lower average intensity of the running difference images during solar minimum and a lower image cadence (30 min 1998 versus 23 minutes in 2000). 
All parameters derived from coronagraphic data  are subject to severe projection effects that results in systematic inaccuracies. A study by \cite{2004JGRA..10903103B} on a set of 111 limb CMEs identified in SMM data gives estimations for `true' values of the CME parameters.

\subsection{Detection of first appearance}\label{1}

\begin{figure}[h]\centering
\includegraphics[width=\columnwidth]{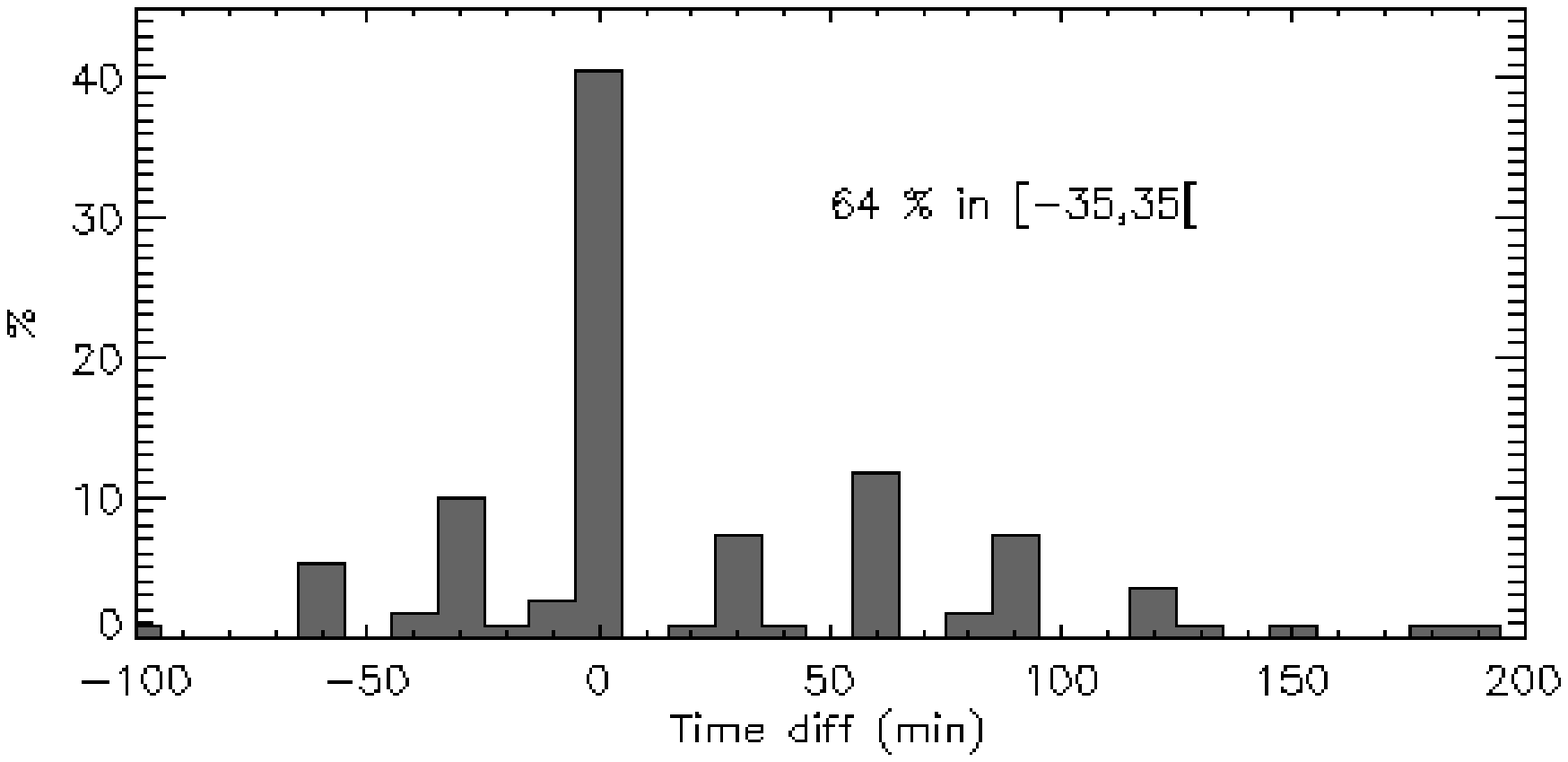}
\includegraphics[width=\columnwidth]{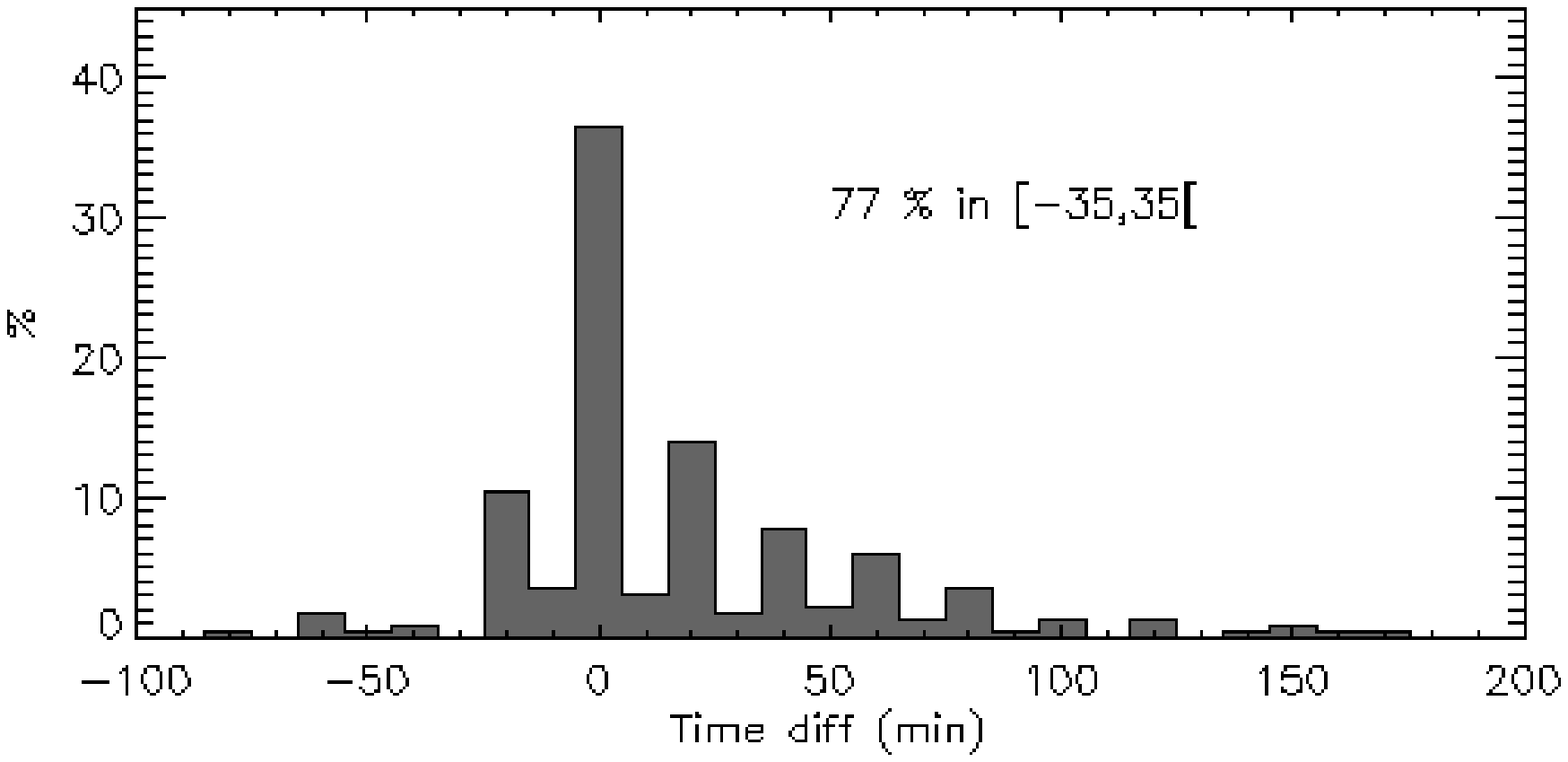}
\caption{Histogram of CDAW-CACTus time differences of first
detection for the 1998 sample (top) and the 2000 sample (bottom).
The binsize was set to 10 minutes. The histogram is heavily biased
by the time-spacing between the LASCO C2-images. The nominal time
spacing peaks around 23 minutes in 2000, and around 30 minutes
in 1998.\label{fig:comp-t}}
\end{figure}

\placefigure{fig:blowout}
Figure \ref{fig:comp-t} shows a histogram of time differences (CDAW-CACTus) of first detection. The binsize was set to 10 minutes. The histogram is heavily biased by the time-spacing between the LASCO C2-images ($\sim 23$ minutes in 2000 and $\sim 30$ minutes in 1998). From the histogram we deduce that during solar max, 77 \% of the first detections differed maximal 1 image and during solar min the corresponding number is 64 \%. This is a good result given the fact that CACTus approximates the CME trajectory linearly. Both physical and technical reasons account for a difference in detection of first appearance (both earlier and later), we list some of them below:

\begin{figure}[h]\centering
\includegraphics[width=\columnwidth]{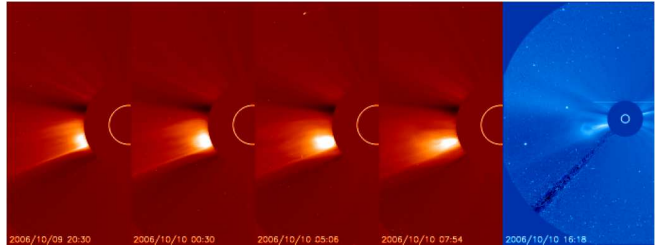}
\caption{The first appearance of a CME is not always well defined.
Here we show a sequence of C2 and one C3 background subtracted images. A CME is seen erupting from the east limb, the
CACTus speed measured was around 150 km/s. It is impossible to
distinguish the background intensity prior to the event (streamer)
and the erupting CME. \label{fig:blowout}}\end{figure}

[1] CMEs can drive waves or shocks ahead of them \citep[e.g.][]{2003ApJ...598.1392V}. They can be observed as a bright (but faint) area prior to the bulk CME eruption. 

[2] \cite{1978SoPh...60..155J} have observed `forerunners', which are described as regions where the corona is slightly more dense than its pre-transient state. In LASCO data we could also observe several cases where a slow rise in intensity is seen before the actual CME is observed. Depending on its intensity it will be detected as `first appearance of the CME' by the observer/detection scheme. 

[3] Another underlying mechanism causing a not sharp transition in intensity from background to CME is the pre-existence of bright material or the very slow rise of a bright structure, prior to the eruption. This is typically the case, for the so-called `streamer blowouts' in which the streamer material is blown away as part of the CME (see Fig.~\ref{fig:blowout}).

[4] From Fig.~\ref{fig:comp-t} we can deduce that CACTus has a preference to detect CMEs more often early than late with respect to CDAW. This is a consequence of erroneously linking two sequential detections into one event. This is a typical example where the human interpretation \textit{does} prove to be useful. CACTus detects motion in each (radial) direction independently. Using information on morphology and speed an observer will notice that activity occurring simultaneously comprises of two events. However, even for the observer it is sometimes impossible to decide whether activity distributed around the occulter is actually linked to one another or not.

\subsection{Apparent width of CMEs}\label{2}

The angular width of a CME is a measure of the volume in the corona that is `blown out'. The {\it apparent} width derived from coronagraphic data, indicate the angular size of this volume projected onto the plane of the sky. This angular size, measured as the angular span around the occulter, remains quasi-constant in the C2/C3 FOV while the CME is propagating outwards. This suggests that CMEs expand radially in a self-similar manner \citep{1982ApJ...254..796L,1984ApJ...281..392L} above $2R_{\odot}$. A popular way to envision a CME geometrically is a circular cone \citep{2002JGRA.107hSSH13Z}, having its vertex in the source region on the solar disk and the cone oriented in the direction of CME propagation. In the case of a limb CME, the cone angle corresponds to the angular span measured in projection onto the plane of the sky. The angular width (and latitude) derived from projected images, is only an apparent quantity that depends on the CME orientation with respect to the observer. A CME launched in a direction close to the Sun-Earth direction appears as a `halo' or partial halo around the occulter. In this case the angular width derived from the coronagraphic observation does not have a geometrical meaning. The `cone model' is a simplified picture, measurements of spatial parameters like CME width and latitude are thus only proxies for CME `volume' and radial direction respectively.

\begin{figure*}\centering
\parbox{.49\linewidth}{
\includegraphics[width=\linewidth]{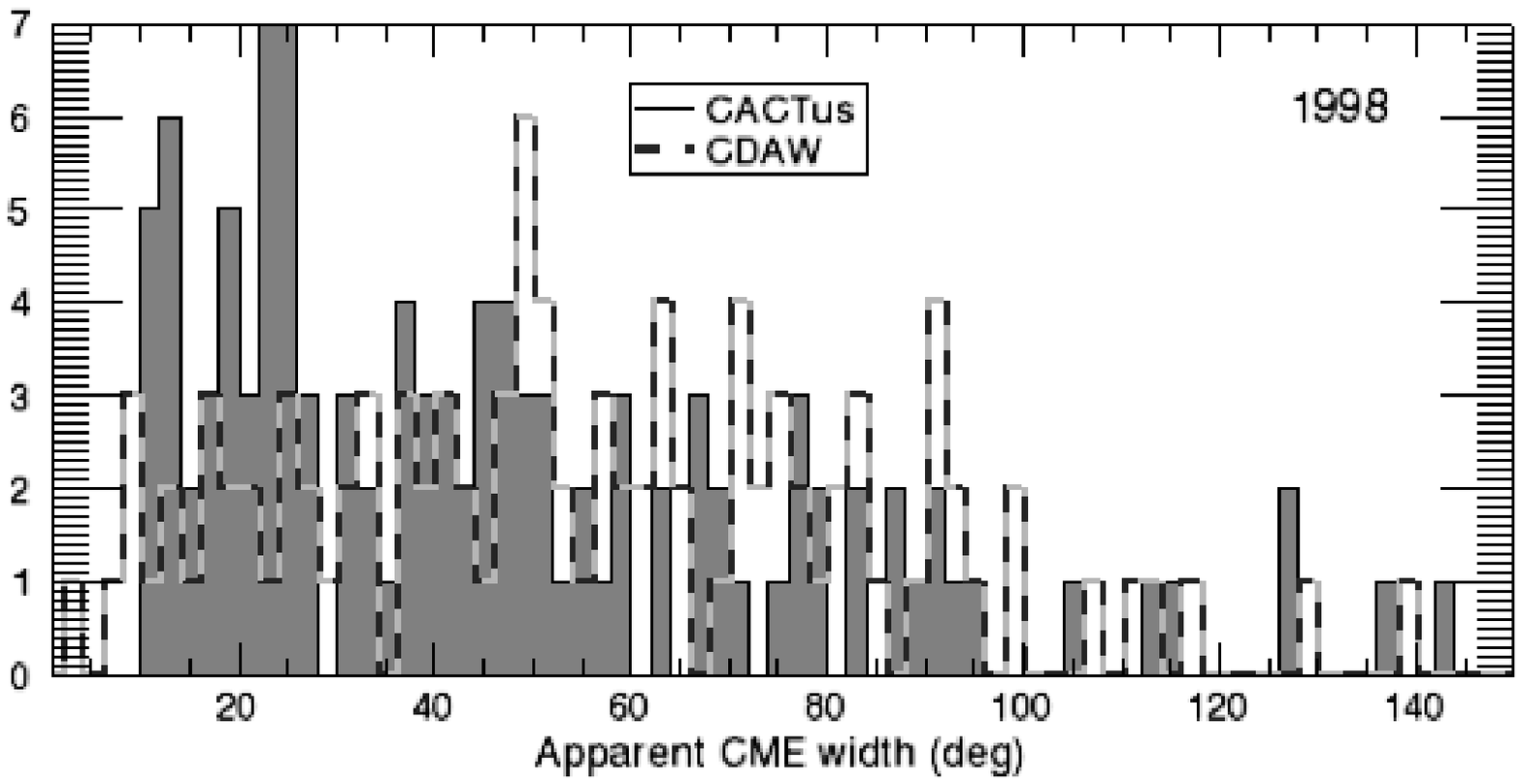}\\
\includegraphics[width=\linewidth]{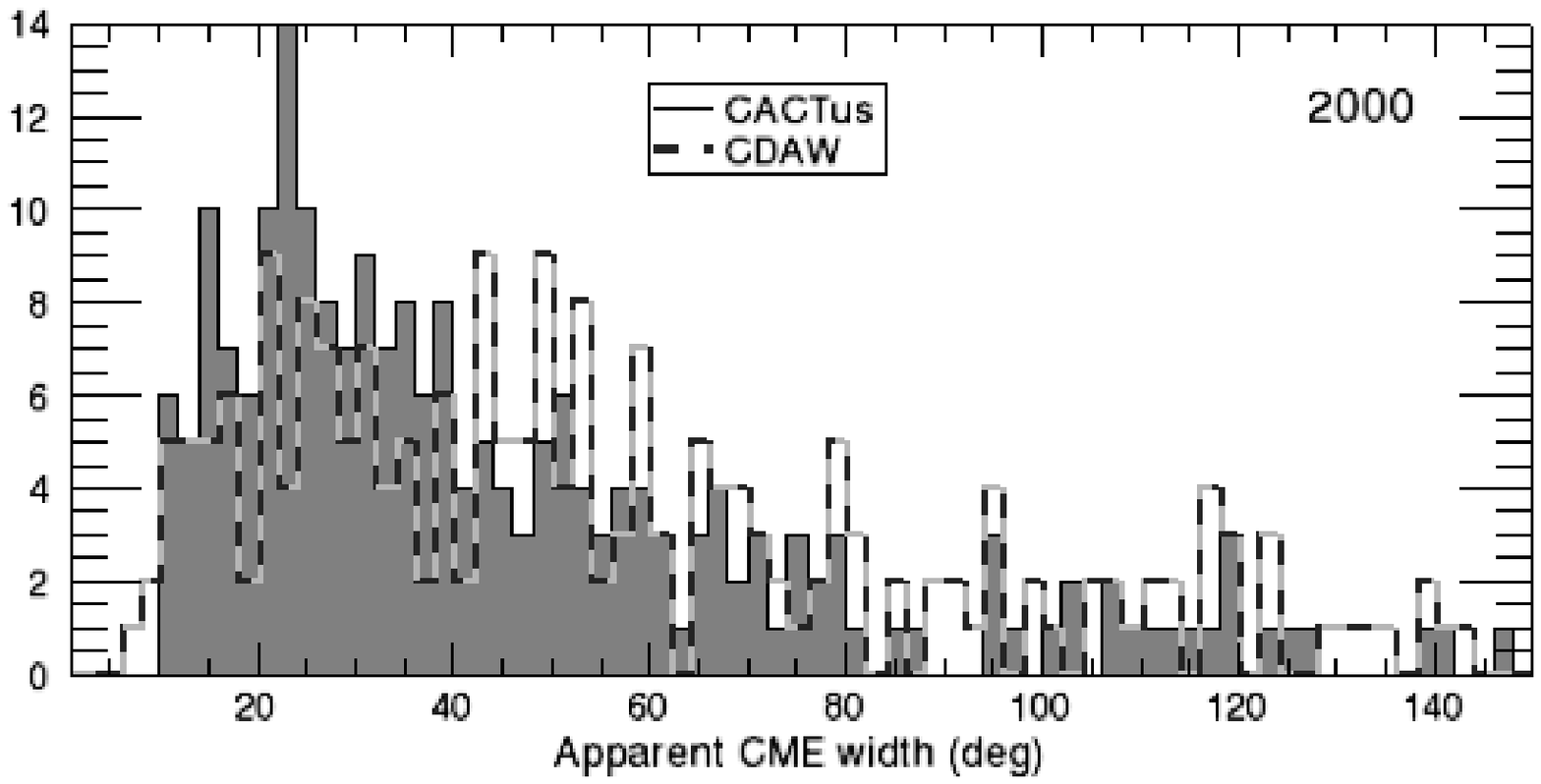}}
\parbox{.49\linewidth}{\includegraphics[width=\linewidth]{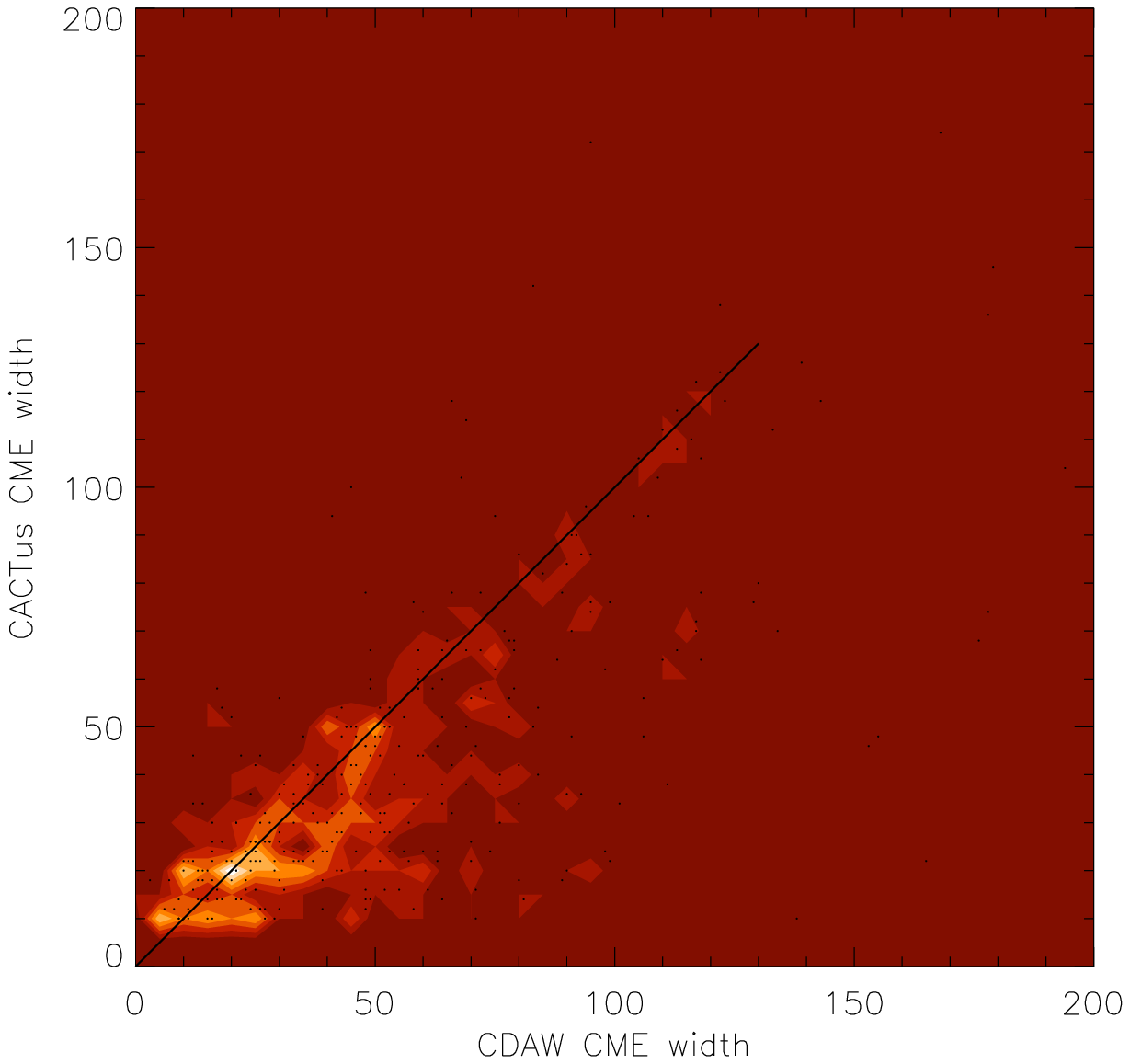}}
\caption{Comparison of the CME widths for the two test-samples.
\textit{Left:} In each graph a histogram of the CME widths is
plotted, with a binsize of 2$^{\circ}$. The upper panel is based on
a sample of 114 CMEs selected in the year 1998 (Solar minimum) and
the lower panel is based on a sample of 222 CMEs selected in the
year 2000 (Solar maximum). As compared to larger statistics
described in this chapter, these histograms appear quite `noisy'.
This is due to the limited sample size. \textit{Right:} Contour plot
illustrating the correspondence between the CDAW and CACTus width
measurements. The line $y=x$ is plotted in black.
\label{fig:comp-da}}\end{figure*}


\begin{figure*}\centering
\includegraphics[width=\linewidth]{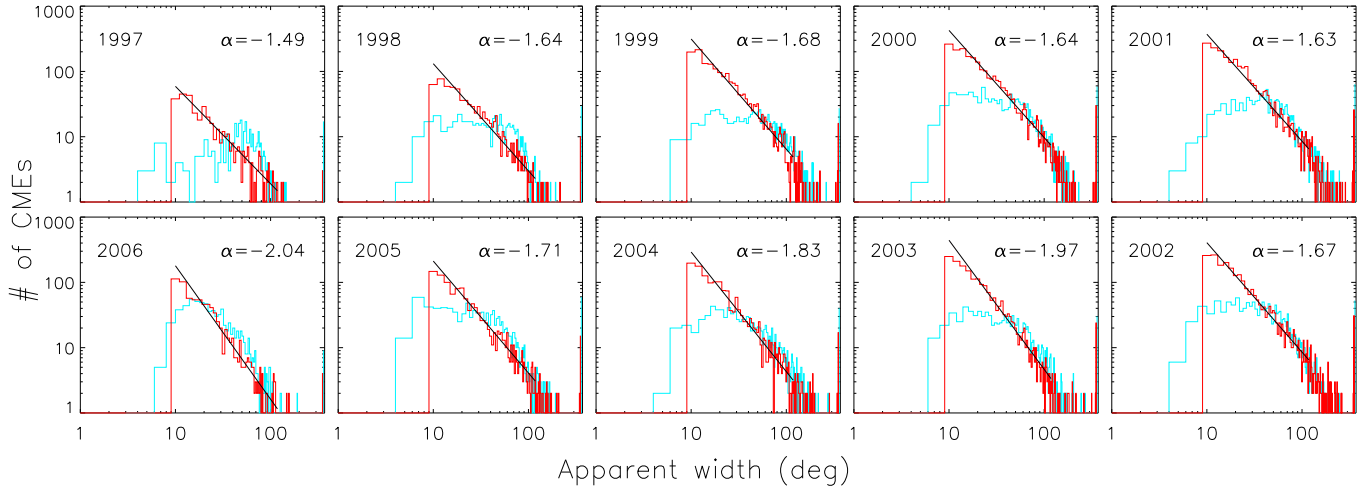}
\caption{Apparent CME width distributions, displayed per year in
log-log scale. The CACTus distribution corresponds to the red
curve, the CDAW distribution is represented by the light blue
curve. The distributions are not corrected for observing time.
\label{fig:width}}
\end{figure*}

\subsubsection{error estimate}
\placefigure{fig:comp-da}
In order to quantify how much the CME width distribution depends on the measurement method we compare the CME widths for the sample of \textit{common} events. CACTus measures the largest width of the CME throughout its outward motion, it is thus not a function of time. In Fig.~\ref{fig:comp-da} (\textit{left}) we have plotted the CME width histograms of the 2 samples in bins of 2\degr, which is the CACTus accuracy. The CACTus width distribution is peaked around 20-25$^{\circ}$. CDAW at the other hand shows a much flatter distribution and measures systematically wider CMEs. At the \textit{right} a contour plot of the CACTus versus CDAW CME widths is shown in the range [0,200]$^{\circ}$. The general direction of the bright contours match well the $y=x$ line. This confirms, at least for events smaller than 120$^{\circ}$, that the CME width indeed \textit{is} a good parameter for estimating the angular size and hence the volume of a CME. However, the large scatter of points indicates that the width is only vaguely defined, and thus space for interpretation is left. For example should CME wave or shock signatures be included when measuring the angular extent of the CME or not? This is not merely a definition issue, the question is rather if an observer is capable to make the distinction between a wave or shock pileup and a `real' CME only based on coronagraphic white light data. A comparative study on `structured CMEs' by  \cite{2004A&A...422..307C} shows indeed that different measurement criteria can lead to substantial differences in CME width measurements (they found differences up to $200\degr$ with values from the CDAW catalog). On average they measured smaller CMEs and less halo's than CDAW, because they did not include deflections of pre-existing structures or shock signatures. Our sample study showed that the CME width is particularly not well defined for CMEs exceeding 120$^{\circ}$, especially halo-CMEs. This is consistent with the result obtained by \cite{2004JGRA..10903103B}, who found a maximum width of $110\degr$ for SMM limb CMEs. Out of the 9 CACTus halo CMEs (from the sample) only 2 of them were also labeled `halo' by CDAW. Inversely CDAW lists 4 halo CMEs which are not labeled halo by CACTus. As a consequence, care has to be taken when interpreting this parameter, especially for large CMEs. 

\subsubsection{CME widths during cycle 23}
\placefigure{fig:width}
The CME width histograms of the two catalogs  are shown in log-log scale in Fig.~\ref{fig:width}. They overlap quite well for CMEs larger than $40\degr$, but show a remarkable difference towards the small side of the `angular spectrum'. The CDAW CME widths are log-normally distributed, broadly peaked around $30\degr$  \citep[e.g.][]{2005ApJ...619..599Y} while the CACTus CME widths could as well suggest a power law behaviour, meaning that the CME widths $\theta$ would be distributed according $ N(\theta)=N_0\theta^{\alpha} \mbox{ with power } \alpha\approx -1.66, $ where $N(\theta)$ is the number of events with angular extent $\theta$ and $N_0$ a constant.  

The question of which distribution provides the best fit to the data (log-normal, power-law) can not be decided solely on the results presented here.  The minimal CACTus CME width was set to $10\degr$, meaning that smaller events were discarded. We therefore do not capture the peak in number of events at small angles - which must exist somewhere - or the rise at even smaller scales. However, the point we wish to stress here is that over a range of scales from 20 to $120\degr$ the CACTus distribution is essentially scale invariant while previous catalogues present a broad maximum around $30\degr$. On the other hand, the scale invariance for events larger than $40\degr$ is consistent for both data sets, shown by the overlap of both curves. In view of Òdescriptive statisticsÓ it is not so important which distribution describes best the data, but seen in perspective of understanding the initiation mechanism and evolution of CMEs, the type of distribution can give hints on the scaling laws that apply to the initiation mechanism. The power law of Fig.~\ref{fig:width} could indicate that eruptions and restructuring of the coronal magnetic field is a scale invariant process: there is no typical size of a CME. For CMEs this would be a new result, but for other types of coronal magnetic field restructuring it is well-known. For flares, for example, \cite{1993SoPh..143..275C} have shown that a power law of $\sim$ -1.6 characterizes the flare energy over 3 orders of magnitude. The fact that exactly the same power law applies for CME widths is intriguing. Probably this is merely coincidence, possibly this hints at common physics of the flare and CME process. 

\begin{figure*}\centering
\includegraphics[width=.48\linewidth]{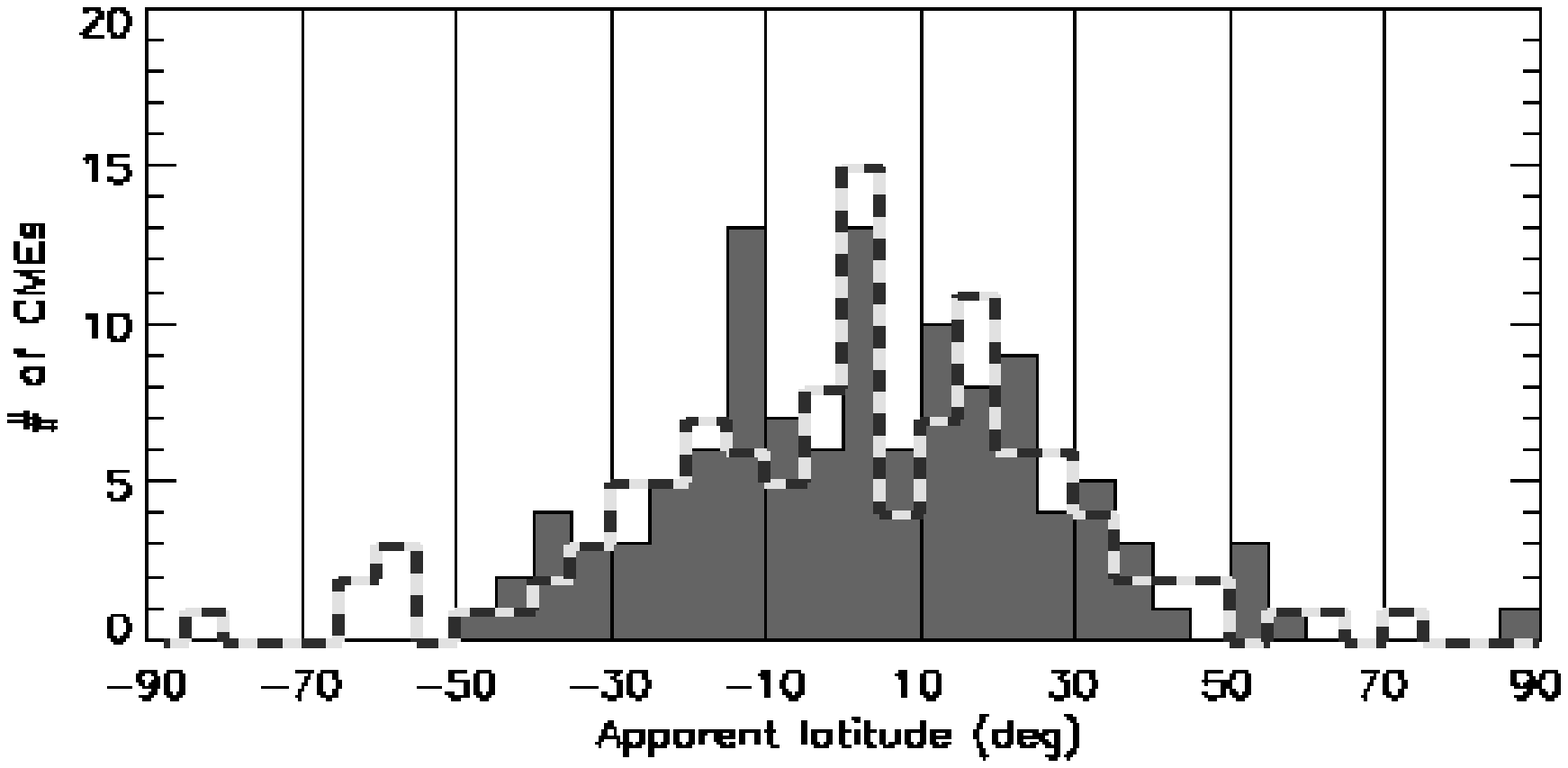}
\includegraphics[width=.48\linewidth]{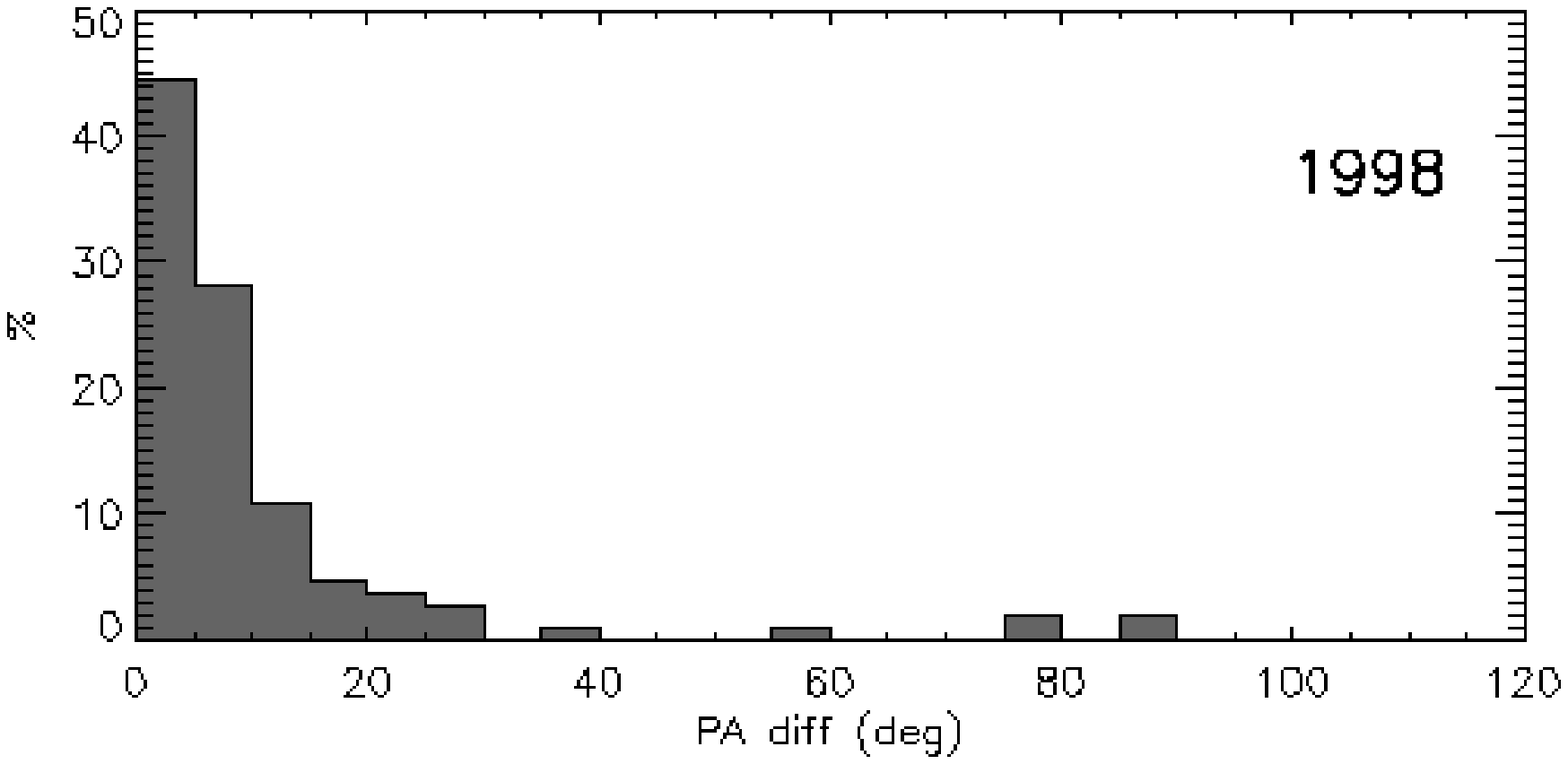}
\includegraphics[width=.48\linewidth]{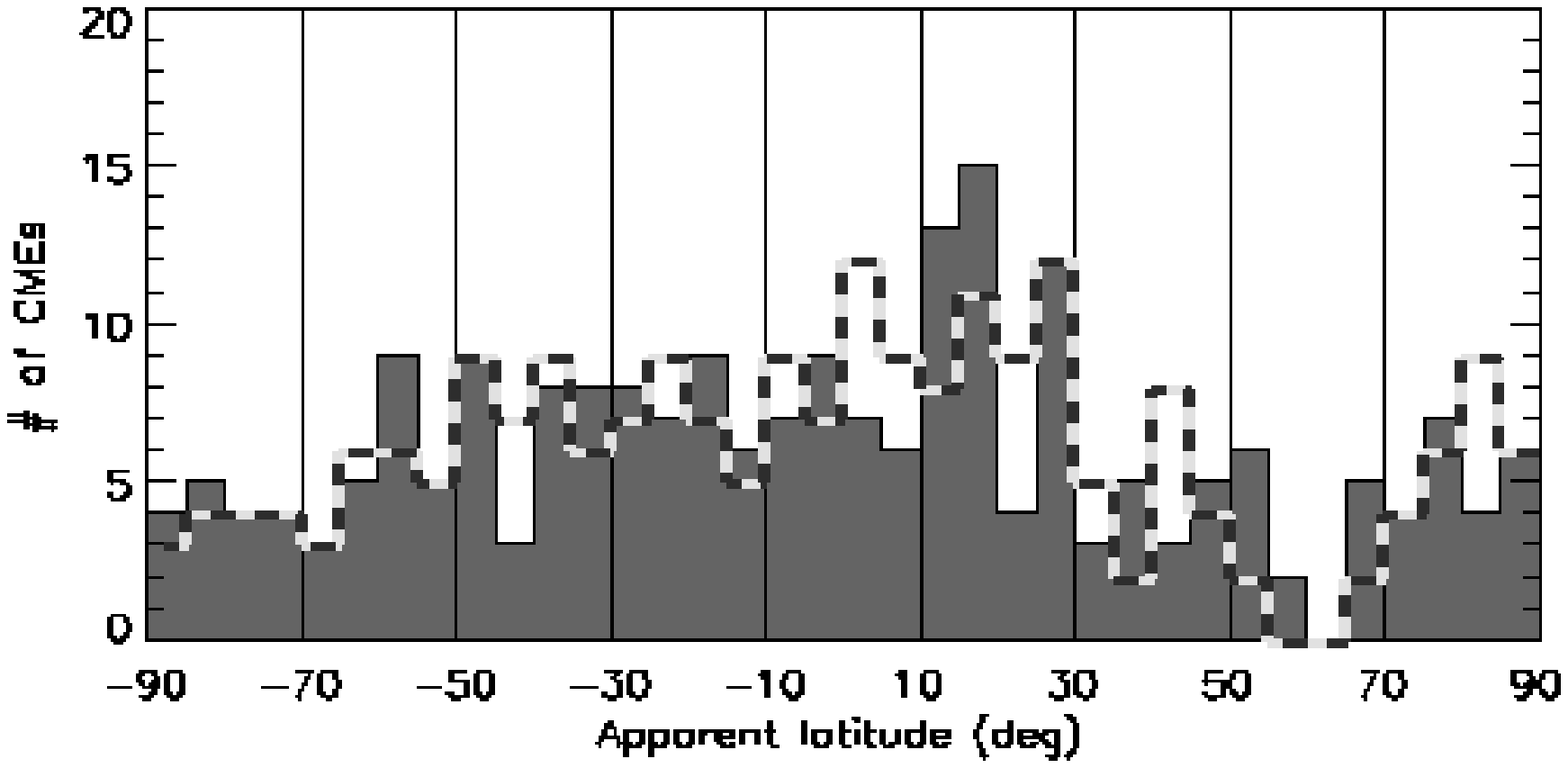}
\includegraphics[width=.48\linewidth]{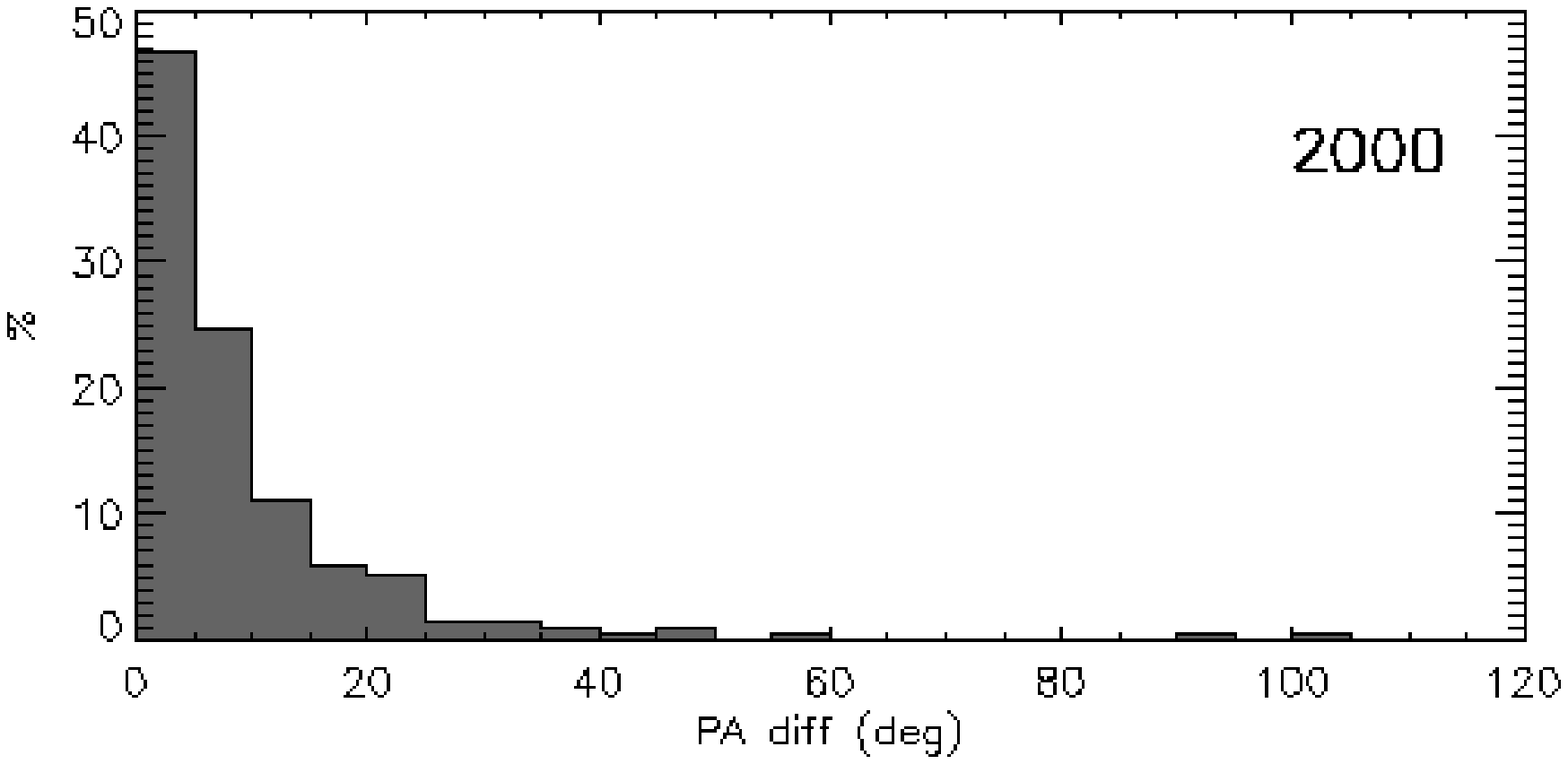}
\caption{\textit{Left:} Sample comparison of the apparent latitude 
distribution. The CACTus latitudes correspond to the filled graph,
while the CDAW latitude distribution is shown in dashed line.
\textit{Right:} Histogram of the difference in principal angle
CACTus-CDAW is plotted, we used a binsize of
5\degr.\label{fig:comp-lat}}
\end{figure*}

\subsection{Apparent latitude of CMEs}\label{3}

The CME {\it projected} latitude is defined as the middle angle of the CME when seen in the white-light images. Due to the projection onto the plane of the sky, projected latitudes are always an upper limit of the true direction of propagation.

\begin{figure}\centering
\includegraphics[width=.7\columnwidth]{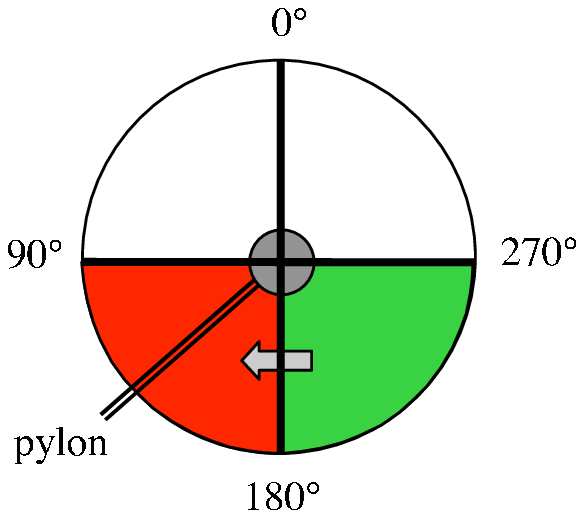}
\caption{Cartoon illustrating the correction we applied to the
latitudinal distributions. The angles indicated are the principal
angles, according to convention running counter clockwise from
solar north. The plot applies when SOHO is in its normal position.
The pylon holding the occulter is then positioned in the
south-east. The histogram in the lower left quadrant is replaced with that
from the lower right quadrant. \label{pylon}}\end{figure}

\subsubsection{error estimate}
\placefigure{fig:comp-lat}
In an attempt to deconvolve the latitudinal distributions from measurement effects, we study the latitudinal differences between the two catalogs, based on our 2 samples of common events. Fig.~\ref{fig:comp-lat} (\textit{right}) shows the histogram of absolute differences in latitudinal measurement. Interpreting these latitudinal differences in terms of measurement uncertainty, we can deduce that measurement errors of (at most) $10\degr$ and $20\degr$ \, apply respectively to 70\% and 90\% of the events for both samples. In the left figure we compare the latitudinal distributions for the two samples (CACTus results corresponds to the filled curve). The only peculiar difference is the peak at $-10\degr$ latitude in the 1998 histogram (\textit{upper left}). We verified the origin of this peak, but did not find a specific reason why CACTus would favour this latitude. All events in this peak, except one, differed less than $20\degr$ from the CDAW value. Hence, we conclude that, the CACTus-CDAW differences in apparent latitude of $\sim 10\degr$ have no significant effect on the latitudinal distributions. {\it Apparent} latitudes seem thus to have small errors introduced by measurement method. We note that latitudes are subject to large projection effects, so care has to be taken when interpreting the results below in terms of {\it true} latitudes. Additionally, CMEs often undergo non-radial motion in the lower corona before they reach the C2 FOV, which makes it difficult to derive CME source-regions from latitudes derived from LASCO C2/C3 observations.

\begin{figure*}\centering
\includegraphics[width=\linewidth]{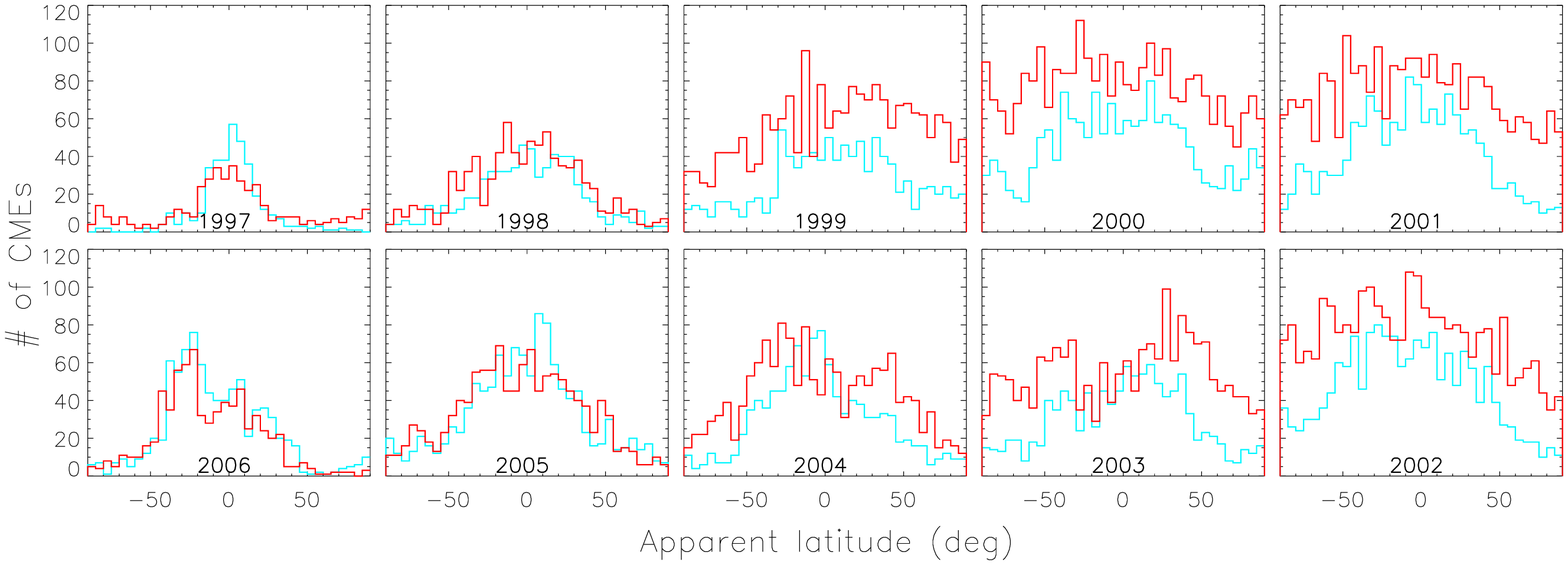}
\includegraphics[width=\linewidth]{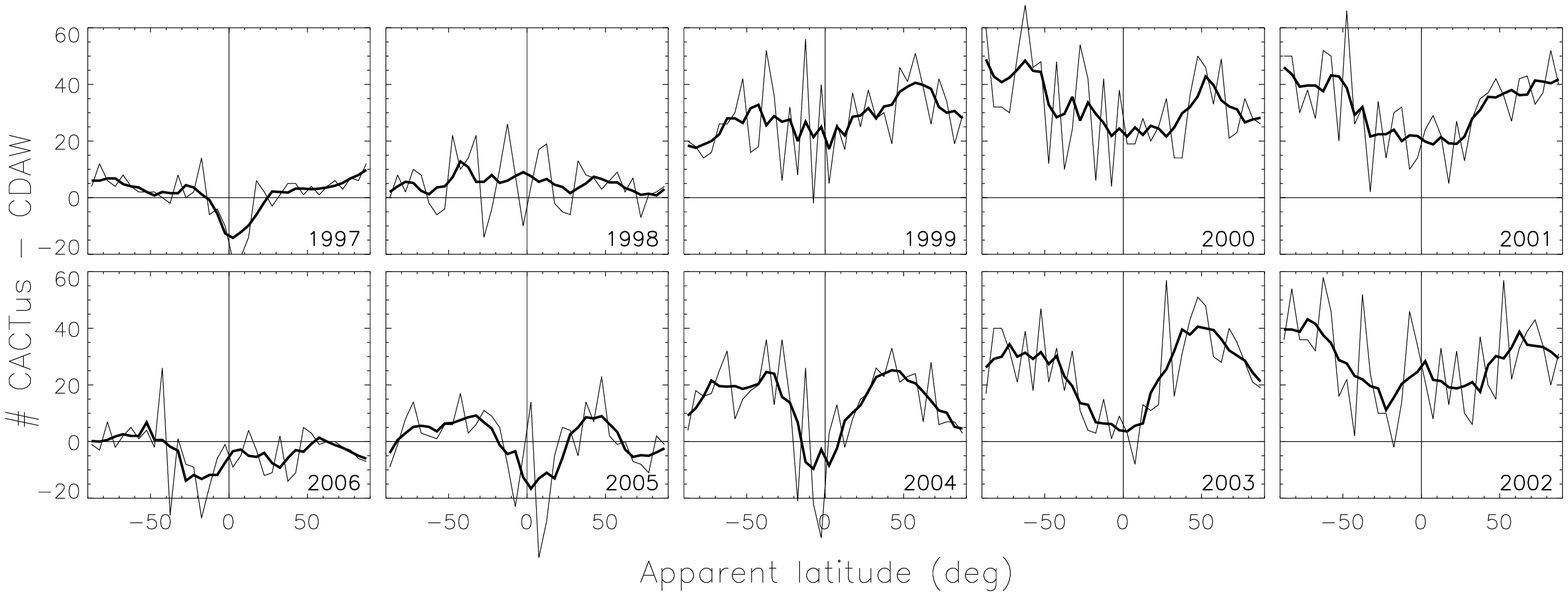}
\caption{\textit{Top:} Yearly histograms of apparent latitudes of
coronal mass ejections. The latitudes run from 0 at the equator to
+/- 90 at the north/south pole. The CACTus distribution
corresponds to the red curve, the CDAW distribution is represented
by the light blue curve. \textit{Bottom:} Difference of latitude
histograms, the thick line is the smoothed curve. Positive values
correspond to more CACTus CMEs.  \label{fig:lat}}\end{figure*}

\subsubsection{CME latitudes during cycle 23}

Fig.~\ref{fig:lat} (\textit{top}) shows the latitudinal
distribution for CACTus (red) and CDAW CMEs (blue) separated for
each calendar year of LASCO observations. The C2 and C3
coronagraphs are both externally occulted. This means a circular
occulting disk is placed in front of the entrance aperture. Hence, no direct
sunlight falls into the instrument, reducing the stray light
significantly. But, as a consequence, the region around the pylon
holding the occulter has a smaller signal to noise ratio. This creates a bias in the latitudinal histograms in the region
around the pylon. To remove this artificial bias
from our statistics, we have corrected the latitudinal
distributions in the direction of the pylon (which is either SE or NW). Assuming that eastern and western
statistics are similar, due to the Sun's rotation, we applied a
correction function to the data, as illustrated in
Fig.~\ref{pylon}. Let $\theta$ be an angle running from $0\degr$ to
90\degr and $N(\theta)$ the number of CMEs with principal angle
$\theta$, then $N(90\degr+\theta) \equiv N(270\degr-\theta)$. We
have applied this correction to the CACTus and CDAW datasets over
the whole period.

\placefigure{fig:lat}
Contrary to what was found for the CMEs angular span, the {\it type}
of the latitudinal distribution does evolve with the solar cycle.
During solar minimum years (1997, 1998 and 2005), the CMEs principal
directions are distributed quasi-normally around the equator in the
range [-20,20]$^{\circ}$. During solar active years, CMEs
erupt almost uniformly at all latitudes, even at higher apparent
latitudes (70$^{\circ}$) in both hemispheres. These findings are consistent with earlier observations
from past cycles \citep{1993JGR....9813177H,1986shtd.symp..107H} and
observations of current cycle \citep{2004shis.conf..201G}. It is important to note that the apparent latitudes are valid for the coronagraphic fields of view after undergoing deflections.  As reported by \cite{2004A&A...422..307C} deflections towards the equator are maximal during solar minimum years due to the presence of the polar coronal holes. The latitudinal CME distribution is thus not only governed by the latitude of the source regions, but also by the presence of coronal holes nearby. 

According to the previous paragraph there is a good correspondence between the global latitudinal properties of CMEs derived from CACTus and CDAW. However, there is a systematic difference. While analyzing the differences in CME width distribution between CACTus and CDAW, we discovered that the systematic higher CME rate, produced by CACTus, is mainly due to small events. Fig.~\ref{fig:lat} (\textit{bottom half}) shows us where these extra events are coming from. In the ascending phase (1998-2002)  the small-scale seems to be randomly distributed. In the descending phase (2003-2005) however, extra events are strongly restricted to two broad bands around $\pm 50\degr$ latitude, bordered by the polar coronal holes at the pole-side and by active regions at the equator-side. No extra events (or even a small deficiency) are observed in the CACTus output \textit{in} the active region band ($<30\degr$). The fact that they are not just randomly distributed, but clearly structured, indicates they are reflecting an underlying large-scale process. This process must be time dependent, or in other words, solar cycle dependent. Further research is required to study this new subpopulation of CME-alikes and their precise source regions on the disc or higher up in the corona.

\subsection{Apparent speed of CMEs}\label{4}

\begin{figure*}[t]\centering
\includegraphics[width=\linewidth]{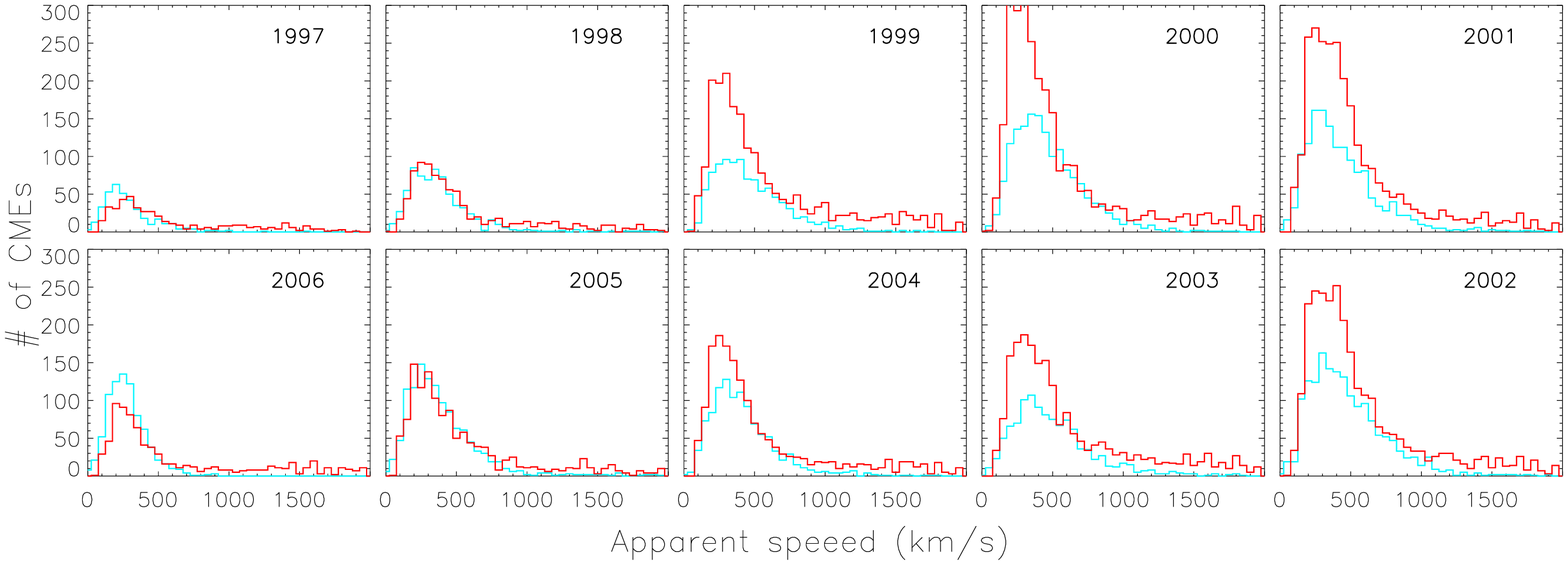}
\caption{Yearly histograms of apparent radial speeds of coronal
mass ejections. The speeds are derived from linear measurements
and do not take into account acceleration or deceleration. We
remind that we can only measure the speed component parallel to
the plane of the sky. The CACTus distribution corresponds to the
red curve, the CDAW distribution is represented by the blue curve.
\label{fig:speed}}
\end{figure*}

Finally, we give an overview of the speed measurements and distributions shown in Fig.~\ref{fig:speed}. The CACTus CME speeds remain lognormally distributed, just like the CDAW speeds
\citep[e.g.][]{2005ApJ...619..599Y}. However, the CACTus CME speed distribution shows a much higher peak, which lies in the range 200-400 km/s.

\subsubsection{error estimate}
\begin{figure*}\centering
\includegraphics[width=.48\linewidth]{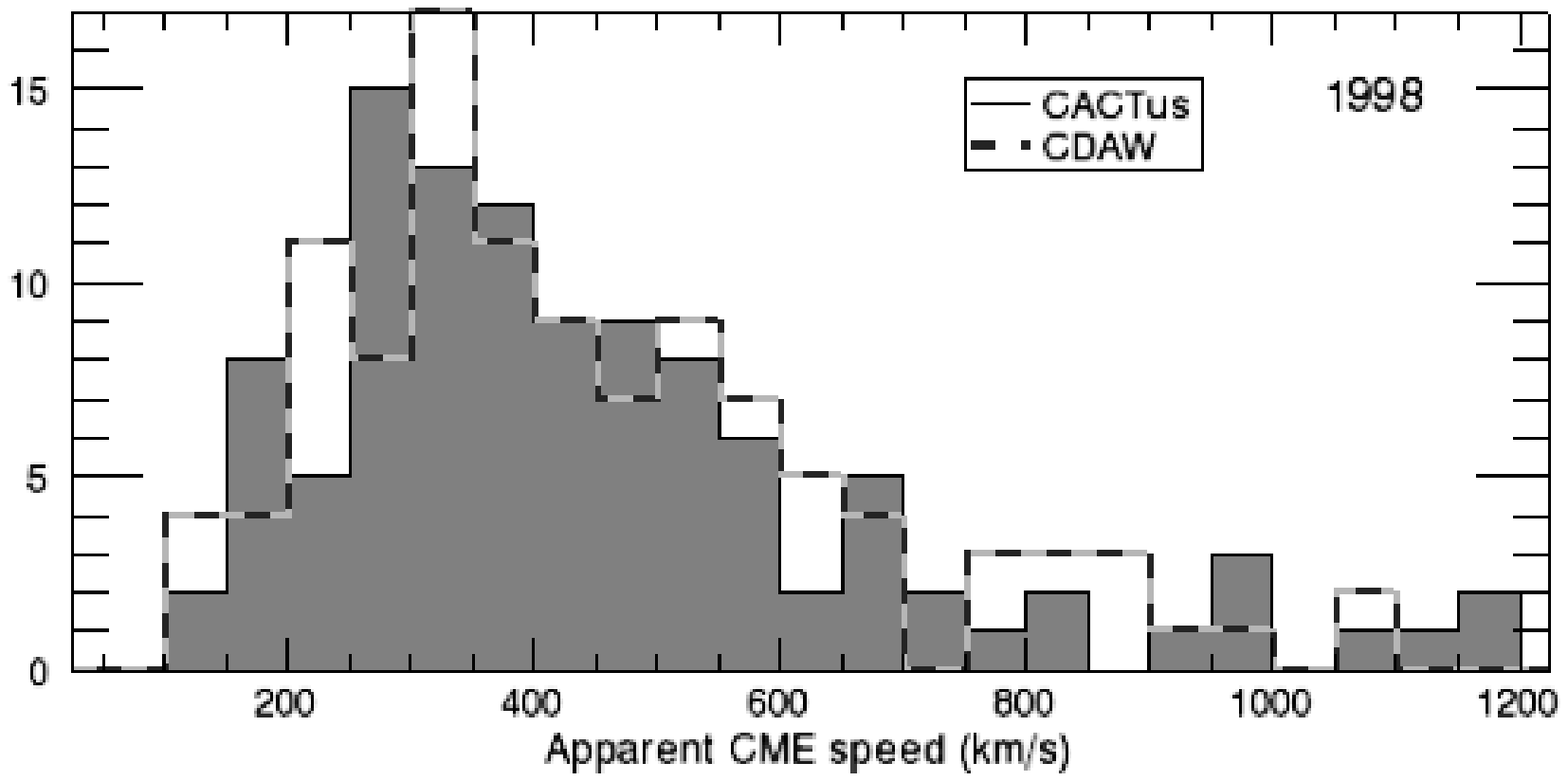}
\includegraphics[width=.48\linewidth]{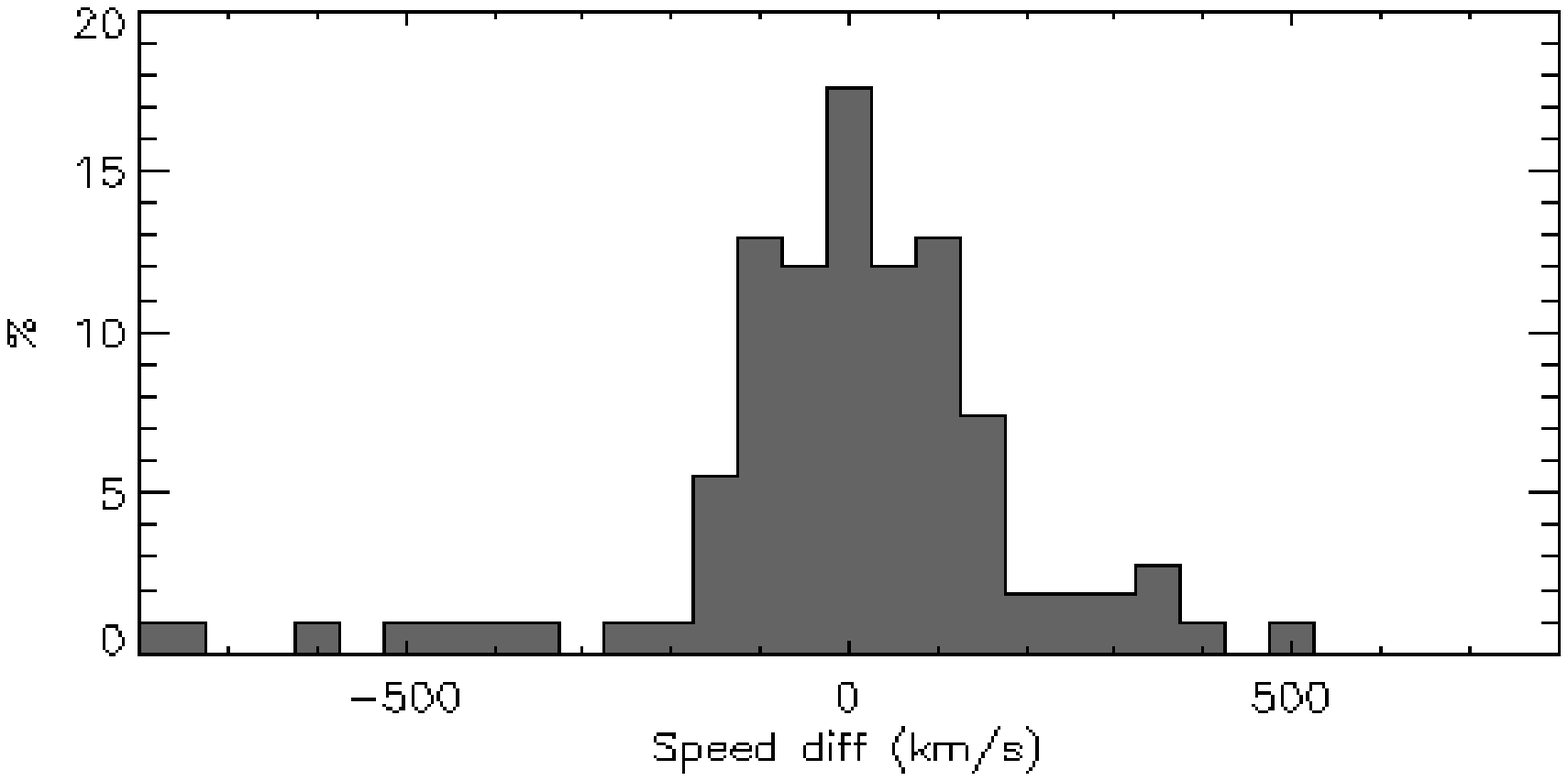}
\includegraphics[width=.48\linewidth]{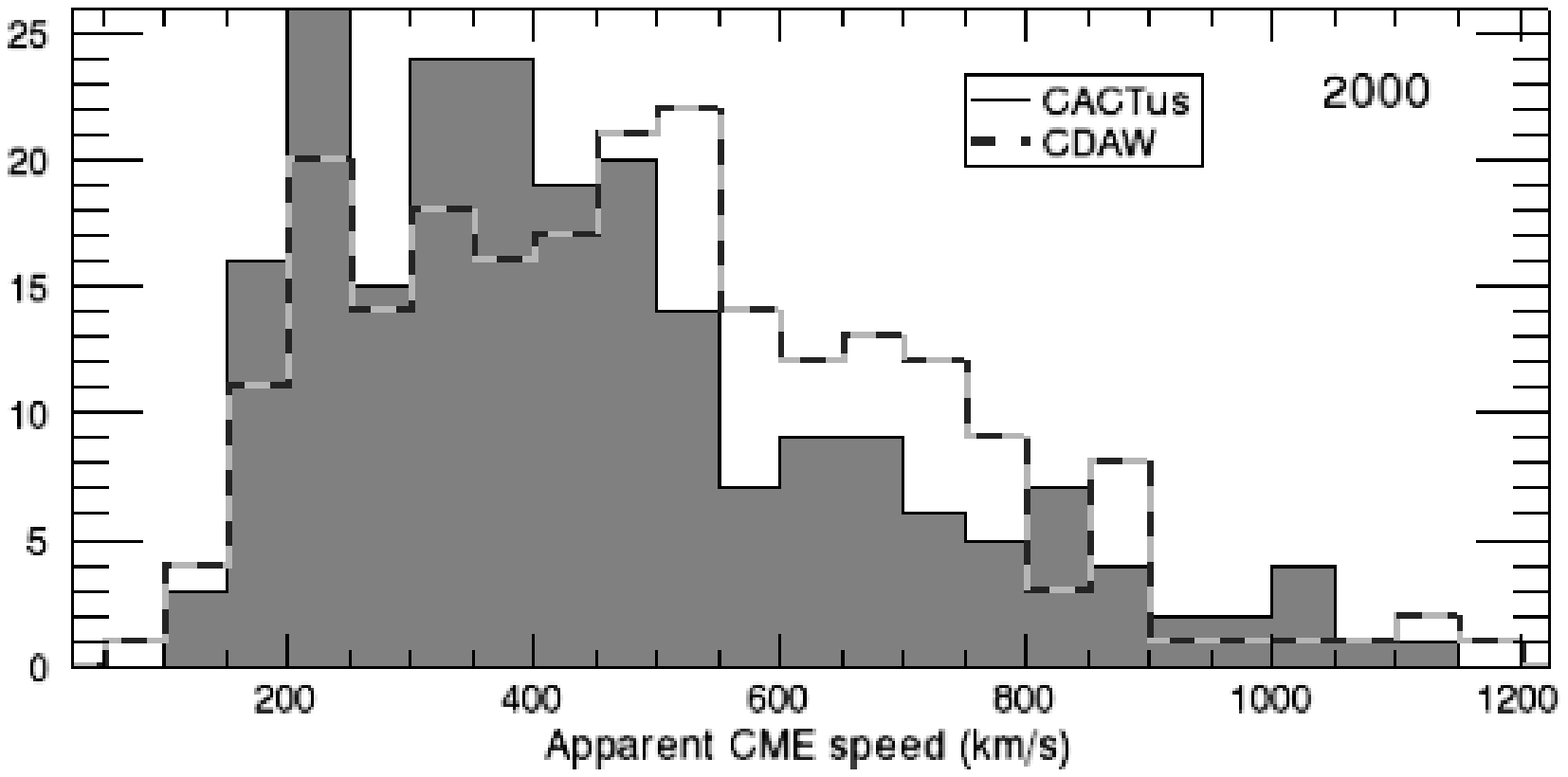}
\includegraphics[width=.48\linewidth]{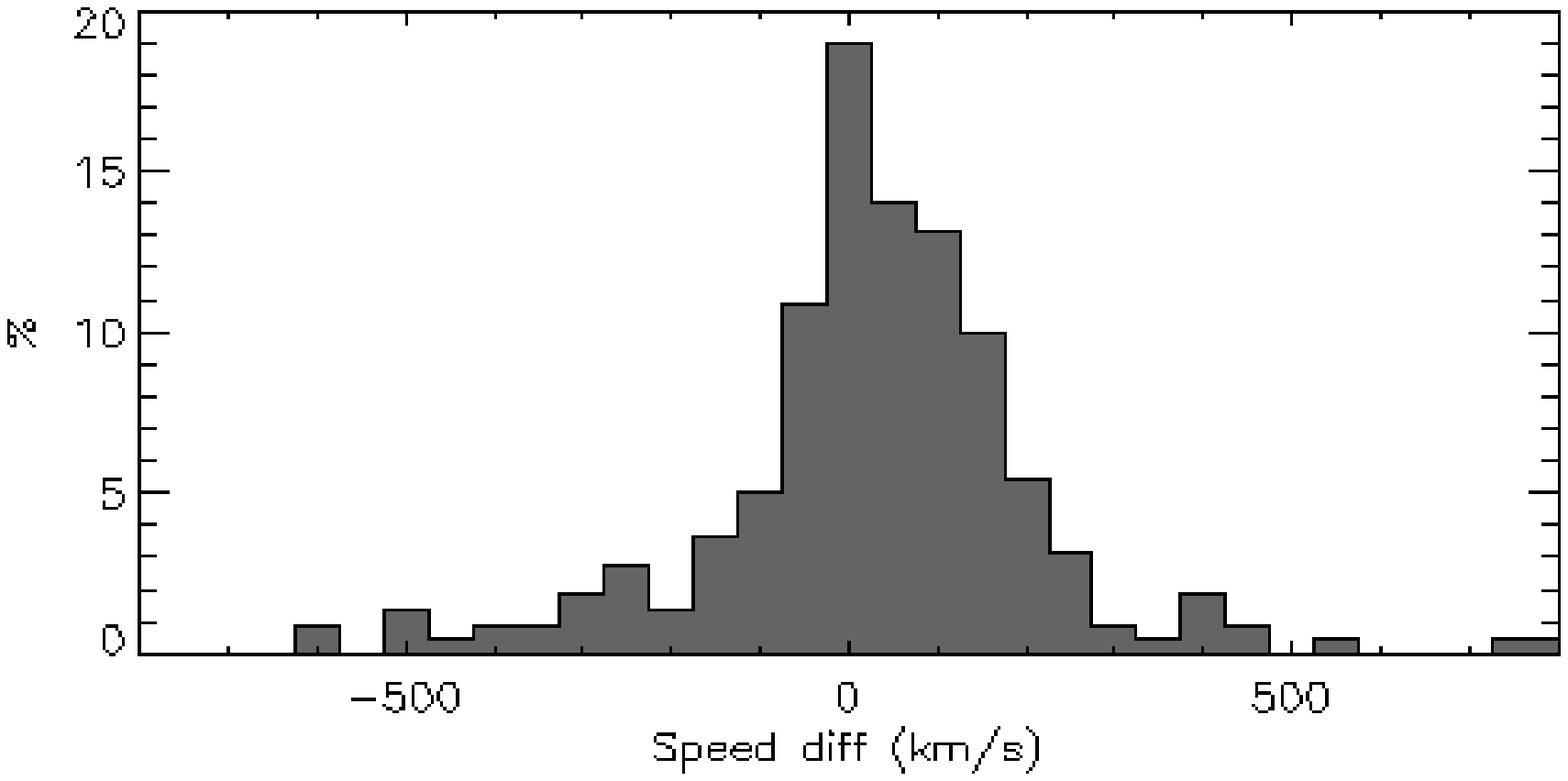}
\caption{\textit{Left:} CME speed distributions compared for both
samples. The CACTus speeds correspond to the filled graph, the
CDAW speed to the dashed line. \textit{Right:} CDAW-CACTus speed
differences. \label{fig:comp-v}}\end{figure*}

\placefigure{fig:comp-v}
CACTus and CDAW speeds differ by definition: CACTus measures a linear speed profile as a function of the angle around the occulter and lists the median value, while the CDAW observer only tracks the fastest moving feature of the leading edge. In this study we compare the CACTus speeds with the linear speeds listed CDAW.  In Fig.~\ref{fig:comp-v} we compare the speed measurement for the two samples of common events. At the left the two histograms are shown and at the right the difference CDAW-CACTus is plotted. For both periods, the difference curve is slightly skewed towards positive difference values inferring a higher CDAW speed is favored. During solar minimum a maximal uncertainty of 175 km/s applies for more than 80\% of the events, during solar maximum the uncertainty is larger.  There are a number of reasons which could contribute to the difference in speed given by CACTus and CDAW:

[1] The majority of large speed-differences occur for narrow CMEs. Possibly, this is because errors on individual measurements are averaged out better for more data points. The CACTus listed speed is the median of all measured speeds in the CME, the more data points, the more reliable this value is.

\begin{figure*}\centering
\includegraphics[width=.48\linewidth]{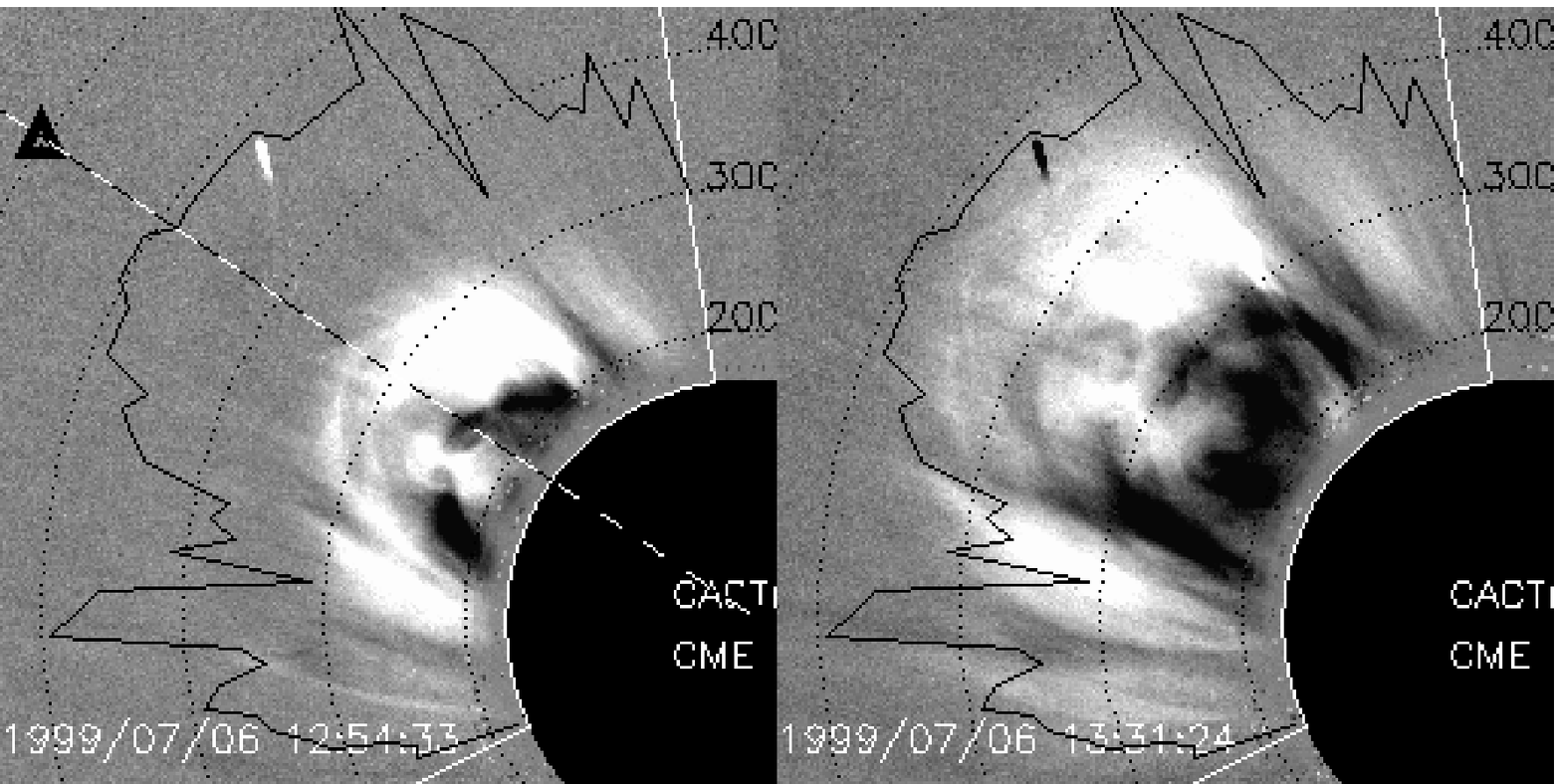}
\includegraphics[width=.48\linewidth]{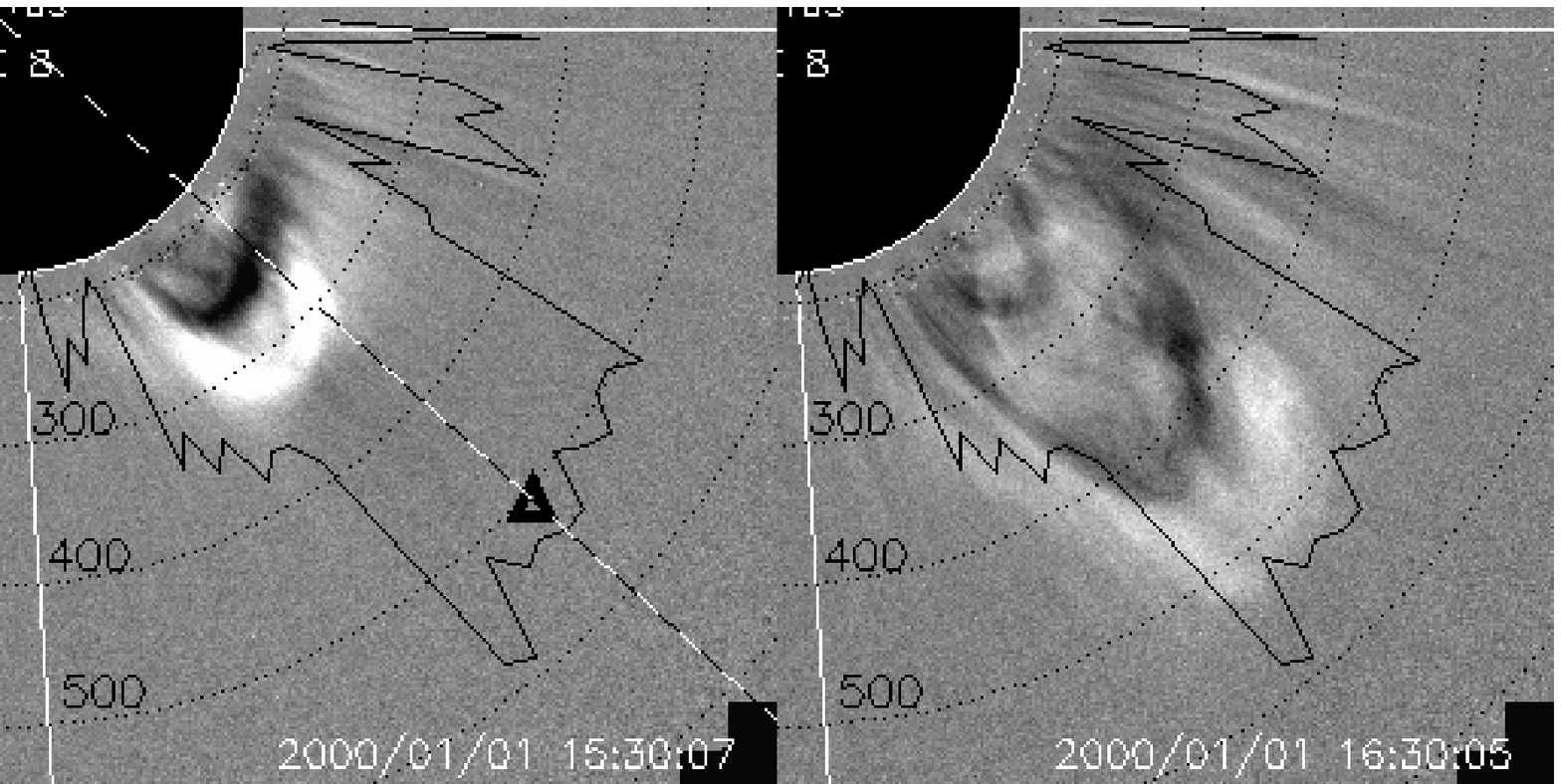}
\includegraphics[width=.48\linewidth]{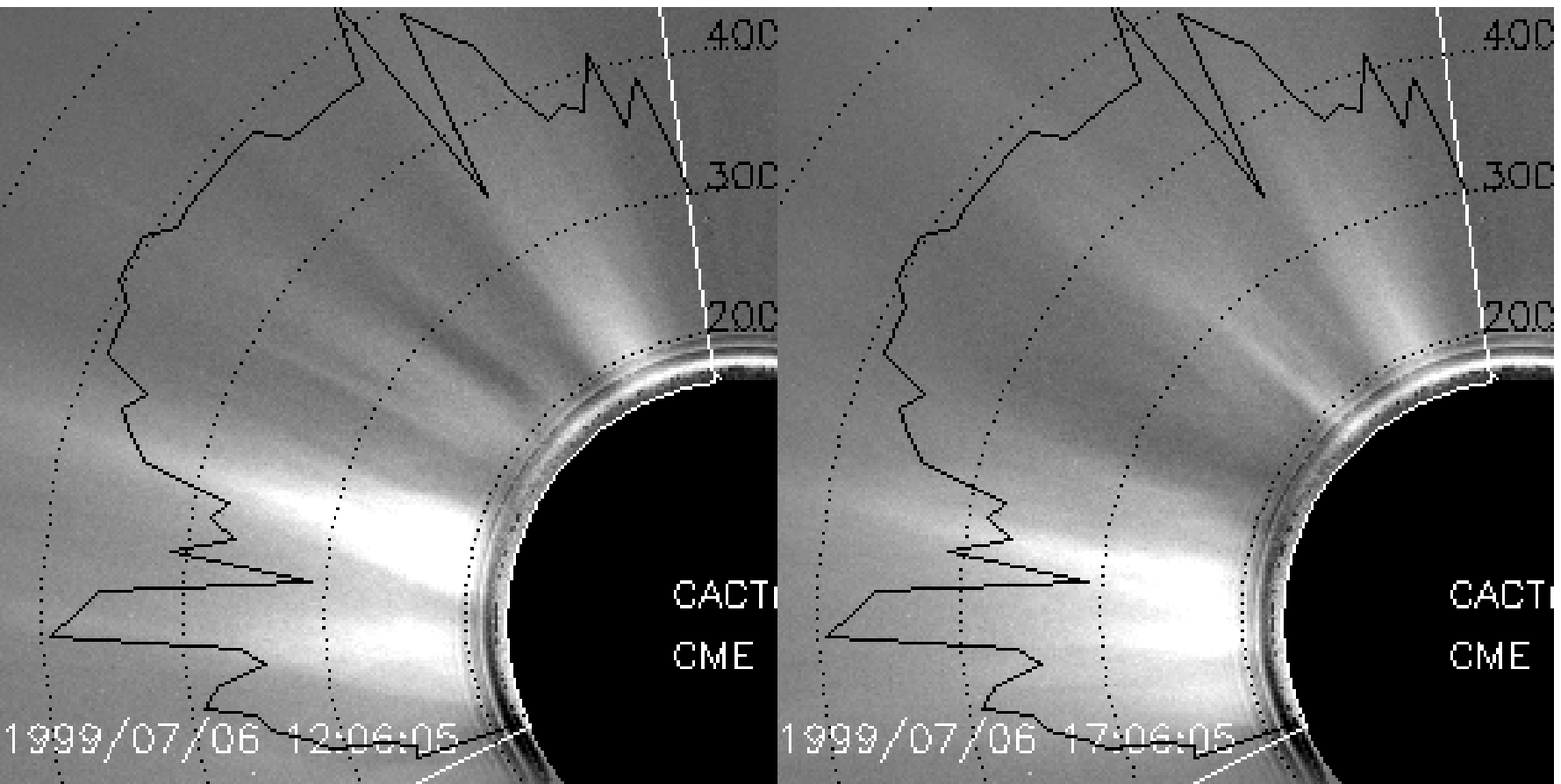}
\includegraphics[width=.48\linewidth]{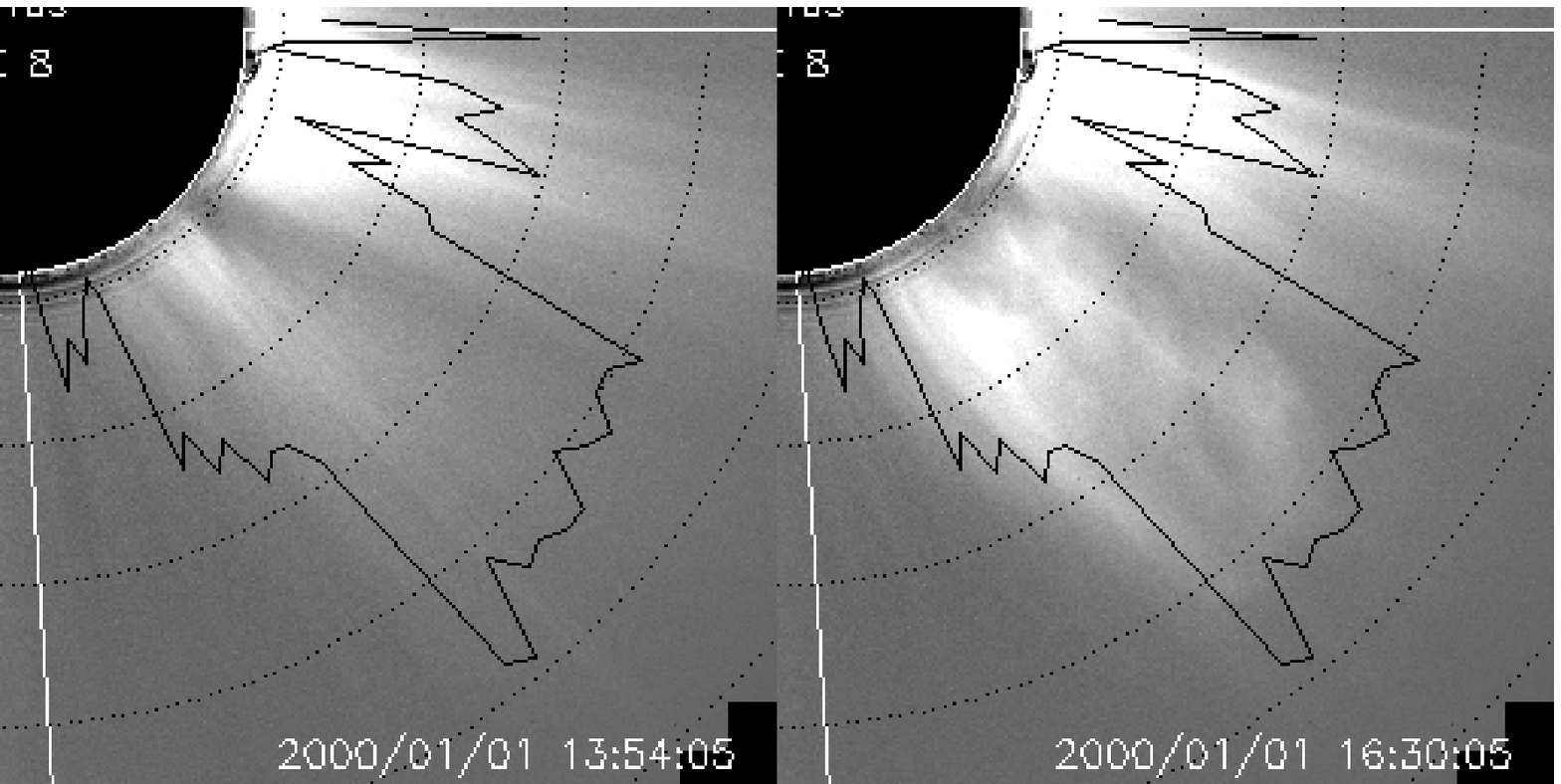}
\caption{Illustration of two limb CME detections in running difference (top) and in background subtracted images (bottom). The speed profile measured by CACTus is shown in black and the CDAW speed is indicated with the black triangle in each top left frame. The velocity scale is indicated in black concentric circles in km/s. {\it Bottom:} For the left event a pre- and a post-CME image are shown to illustrate the streamer displacement. At the right we show a pre-CME image and an image containing the CME. \label{fig:speed1}}\end{figure*}

[2] CMEs have internal speed variability, for example as a consequence of interaction with different background solar wind structures.  As example, we show two limb-CMEs in running difference and their speed-measurement in Fig.~\ref{fig:speed1} (\textit{top}). The CACTus speed profiles are quite uniform at the leading edge. The profiles are both distorted towards the edges of the CMEs. The magnetic and density structure of the ambient corona plays a not unimportant role in the outward evolution of CMEs \citep[e.g.][]{2005A&A...430.1099J} and vice versa. To illustrate this interaction, we have also plotted background subtracted images for these two events. For the first event (left) a pre- and a post-CME image are shown. It can be seen that the brightest streamer is deflected down due to the interaction with the CME. For the second event (right) we show a pre-CME image and an image containing the CME. The helmet streamer at the north was pushed aside during the event, but adapted its original position after the CME had left. The disturbance is traveling outward through the streamer, and the radial component of its speed is captured by CACTus.

[3] For some CMEs there is a large uncertainty on the speed measurement, simply because the `front' of the outward moving feature is not clearly outlined. Even for the rather well observed front of the first CME (erupting to the NE) from Fig.~\ref{fig:speed1} CACTus and CDAW speeds deviate still 100 km/s. 

\section{Discussion on narrow transients}\label{sec:discuss}

A discussion on narrow events, necessarily leads to a discussion on the definition of coronal mass ejections. Many questions arise: Is there a continuum from large coronal mass ejections down to narrow ejections representing the continuous coronal wind outflow? Can we introduce the term micro- or nano-CMEs, cfr.\ nano-flares \citep{1983sowi.conf...23P}? Are narrow ejections a subset of `normal' coronal mass ejections? Or do they form a separate class of events for which the pre-eruptive state is different from the `classical' CME scenario? What can these events, which sometimes occur prior to a larger CME, teach us on the CME initiation mechanism? If CMEs in general contribute to the reorganization of the large-scale magnetic field, does this also apply to these narrow events? If yes, they might act as `lilliputters' gradually untying the magnetic field lines which finally leads to unstable configurations. Too many questions to answer here, and probably several scenarios apply. 

A combination of several criteria makes that some of these events are easily recognized as CMEs and others are not. Observable parameters for CME detection in white-light are: brightness, angular extent, well-defined shape and leading edge, suggestion of magnetic structure, time difference with major events occurring in the same direction.  E.g. jets have an unclear ÔshapeÕ and do not often show an organized structure.  This makes that if the jet is bright and wide, it is included in a catalog, but when it is faint or very narrow (like polar jets), they usually are not included. It seems thus that at least a number of the above criteria has to be fulfilled in order to Ôcount themÕ as CMEs.  In a close inspection of 171 `CACTus only' events (from the sample in 1998 and 2000), we find that the majority occur during times of other activity. Here below is a list that tries to describe the different types of small events that we encountered:
\begin{itemize}
\item[-] events split in space or time from another event, it is often not clear if there is an actual physical connection between the events or if they are just causally related. (32, 18.7\%)
\item[-] trailing outflow from the CME footpoints  (28, 16.3\%)
\item[-] general activity, may be during or after a large CME (26, 15.2\%)
\item[-] stand alone events, including jets and recurrent events, sometimes ahead of a large CME (25, 14.6\%)
\item[-] wave-like disturbances traveling through dense regions (e.g. streamers) (19, 11.1\%)
\item[-] false detections (13, 7.6\%)
\item[-] unclear faint detections (12, 7.0\%)
\item[-] slowly rising loop-like structures, typically during the evolution of a streamer blowout. (10, 5.8\%)
\item[-] opening field lines that are crossing have a higher intensity and result in an apparent `blob' moving outward (5, 2.9\%) 
\end{itemize}

The above list shows that narrow/small events, do not form one separate category, but have a variety in physical appearance.  About 60\% of the CACTus-only events are related to a larger eruption or reflect the high degree of activity in the corona (bursty outflow during solar active times).  A small fraction (15\%) are independent events that do not show any direct link to a large CME, e.g. jets \citep{2002ApJ...575..542W}. Also, halo CMEs are often not recognized as such by CACTus, because the interconnection between outflow in different directions may be too faint to be detected, and this results in several smaller detections. The majority of narrow events occur thus as a sign of high coronal activity, i.e. in conjunction with well established CMEs.

The statistics based on the CACTus observations leads to the idea that a coronal mass ejection is not an `atomic' process, but a sequence of mass expulsions of which the dominant one is generally recognized as a (flux-rope) CME. The bursty small-scale outflow, observed prior to, simultaneously with, or in the aftermath of the dominant eruption is interpreted as being the result of multiple reconnections.  This hints towards the existence of multiple thin current sheets over the total volume of the eruption, rather than a single monolithic current sheet.  Ample observational and numerical evidence proves their existence and dynamics observed as bursty outflow \citep[e.g.][and references therein]{2003ApJ...594.1068K,2003JGRA..108.1440W,2007ApJ...655..591R,2002ApJ...575.1116C,2006ApJ...638.1110B}. The post-eruptive blobs seem to be similar to the blobs observed by \cite{1997ApJ...484..472S} in streamer stalks. 

A cascade of events down to smaller scales is the typical characteristic of self organized systems and avalanche models. The observed power-law in the CACTus CME width distribution suggests that coronal mass outflow is scale invariant,  at least in the range of scales from 20 to $120 \degr$. The application of scale invariance to processes in the solar corona is extensively studied for solar flares \citep{1991ApJ...380L..89L,1993SoPh..143..275C} and is also investigated for the acceleration of high energetic particles \citep{2004ApJ...608..540V,2006SSRv..124..249C}. Since all three processes are the result of rapid release of magnetic energy, it would be not surprising that scale invariance also applies to the CME eruption process.  Specific studies on the mechanisms governing CME eruptions of various sizes are required to further this interpretation. 

\section{Conclusion}

In this paper we compare our statistics and measurements of CME parameters against the CDAW LASCO CME catalog  \citep{2003iscs.symp..403G, 2004JGRA..10907105Y} as reference catalog. In this catalog small events like jets (not polar jets) are generally listed when they are distinct and bright enough. For the majority of large well defined events there is a relatively good agreement, but there are also periods where the two catalogs do not agree at all. This is because at some times, coronal activity is omni-present, faint and unstructured for example while the corona is restructuring hours to days after a large CME has erupted. 

The CACTus CME rate follows the solar cycle, and changes roughly with a factor 4 between minimum and maximum. After applying a correction factor we find that the CACTus CME rate is surprisingly consistent with CME rates found during the past 30 years. The CME rate shows a delay of 6 to 12 months with respect to the sunspot index. The CACTus CME rate decreases in the descending phase whereas the CDAW CME rate remains quasi-constant between 2004 and 2007, probably due to changes in observation criteria adopted by the operator. CME width and speed distributions do not show a great variation over the solar cycle, whereas the latitude histograms evolve from Gaussian during solar minimum, to multimodal during solar maximum years, showing that coronal mass is erupting at all {\it projected} latitudes.

A comparison of a sample of common events shows that the CDAW CME width is on average wider than the CACTus CME width. Confusion in this parameter exists, since it is difficult to disentangle plasma from wave and shock signatures. Streamer deflections are generally not included in a CME width measurement, but bright waves or shocks sometimes are included by the operator whereas CACTus only applies a brightness criterion. There is a particularly bad overlap in halo CMEs, because they usually have several parts of lower intensity. The latitude measurements are quite compatible between CDAW and CACTus with 70\% of the events having a difference below $10\degr$ in latitude. More than 80\% of the CACTus-CDAW speeds differences are in the $\pm$ 175 km/s range.  

Our statistics show that small scale outflow is ubiquitously observed in white light data. Overall, CACTus detects many more events than CDAW, because it tracks all outward moving features. A sample study of CACTus-only events shows that the majority (about 60\%) are small events related to previous CMEs or to high coronal activity (bursty outflow). Also individual events were detected (about 15\%), thus small events are not a mere by-product of large well-established CMEs. The CACTus and CDAW CME width distributions diverge significantly for widths smaller than $40\degr$. The CACTus distribution is essentially scale invariant over a range of scales from 20 to $120\degr$. This supports the hypothesis that the corona indeed is a self-organized system, an idea that has been developed in relation to the scale-invariance of flares and the acceleration of particles.

\acknowledgments
\textbf{Acknowledgments.} ER would like to thank Russ Howard for insightful discussions and Spiros Patsourakos for useful comments. This work is supported by PRODEX contract C90204 (Solar Drivers of Space Weather),  managed by the European Space Agency in collaboration with the Belgian Federal Science Policy Office.

\bibliographystyle{apj}

\begin{thebibliography}{46}
\expandafter\ifx\csname natexlab\endcsname\relax\def\natexlab#1{#1}\fi

\bibitem[{{Bachmann} \& {White}(1994)}]{1994SoPh..150..347B}
{Bachmann}, K.~T. \& {White}, O.~R. 1994, \solphys, 150, 347

\bibitem[{{Bemporad} {et~al.}(2006){Bemporad}, {Poletto}, {Suess}, {Ko},
  {Schwadron}, {Elliott}, \& {Raymond}}]{2006ApJ...638.1110B}
{Bemporad}, A., {Poletto}, G., {Suess}, S.~T., {Ko}, Y.-K., {Schwadron}, N.~A.,
  {Elliott}, H.~A., \& {Raymond}, J.~C. 2006, \apj, 638, 1110

\bibitem[{{Berghmans}(2002)}]{2002svco.conf...85B}
{Berghmans}, D. 2002, in ESA SP-506: Solar Variability: From Core to Outer
  Frontiers, ed. J.~{Kuijpers}, 85--89

\bibitem[{{Brueckner} {et~al.}(1995){Brueckner}, {Howard}, {Koomen},
  {Korendyke}, {Michels}, {Moses}, {Socker}, {Dere}, {Lamy}, {Llebaria},
  {Bout}, {Schwenn}, {Simnett}, {Bedford}, \& {Eyles}}]{1995SoPh..162..357B}
{Brueckner}, G.~E., {Howard}, R.~A., {Koomen}, M.~J., {Korendyke}, C.~M.,
  {Michels}, D.~J., {Moses}, J.~D., {Socker}, D.~G., {Dere}, K.~P., {Lamy},
  P.~L., {Llebaria}, A., {Bout}, M.~V., {Schwenn}, R., {Simnett}, G.~M.,
  {Bedford}, D.~K., \& {Eyles}, C.~J. 1995, \solphys, 162, 357

\bibitem[{{Burkepile} {et~al.}(2004){Burkepile}, {Hundhausen}, {Stanger},
  {St.~Cyr}, \& {Seiden}}]{2004JGRA..10903103B}
{Burkepile}, J.~T., {Hundhausen}, A.~J., {Stanger}, A.~L., {St.~Cyr}, O.~C., \&
  {Seiden}, J.~A. 2004, Journal of Geophysical Research (Space Physics), 109,
  3103

\bibitem[{{Cargill} {et~al.}(2006){Cargill}, {Vlahos}, {Turkmani}, {Galsgaard},
  \& {Isliker}}]{2006SSRv..124..249C}
{Cargill}, P.~J., {Vlahos}, L., {Turkmani}, R., {Galsgaard}, K., \& {Isliker},
  H. 2006, Space Science Reviews, 124, 249

\bibitem[{{Ciaravella} {et~al.}(2002){Ciaravella}, {Raymond}, {Li}, {Reiser},
  {Gardner}, {Ko}, \& {Fineschi}}]{2002ApJ...575.1116C}
{Ciaravella}, A., {Raymond}, J.~C., {Li}, J., {Reiser}, P., {Gardner}, L.~D.,
  {Ko}, Y.-K., \& {Fineschi}, S. 2002, \apj, 575, 1116

\bibitem[{{Cremades} \& {Bothmer}(2004)}]{2004A&A...422..307C}
{Cremades}, H. \& {Bothmer}, V. 2004, \aap, 422, 307

\bibitem[{{Cremades} \& {St.~Cyr}(2007)}]{2007AdSpR..40.1042C}
{Cremades}, H. \& {St.~Cyr}, O.~C. 2007, Advances in Space Research, 40, 1042

\bibitem[{{Crosby} {et~al.}(1993){Crosby}, {Aschwanden}, \&
  {Dennis}}]{1993SoPh..143..275C}
{Crosby}, N.~B., {Aschwanden}, M.~J., \& {Dennis}, B.~R. 1993, \solphys, 143,
  275

\bibitem[{{Donnelly} {et~al.}(1983){Donnelly}, {Heath}, {Lean}, \&
  {Rottman}}]{1983JGR....88.9883D}
{Donnelly}, R.~F., {Heath}, D.~F., {Lean}, J.~L., \& {Rottman}, G.~J. 1983,
  \jgr, 88, 9883

\bibitem[{{Gnevyshev}(1967)}]{1967SoPh....1..107G}
{Gnevyshev}, M.~N. 1967, \solphys, 1, 107

\bibitem[{{Gopalswamy}(2004)}]{2004shis.conf..201G}
{Gopalswamy}, N. 2004, in ASSL Vol. 317: The Sun and the Heliosphere as an
  Integrated System, ed. G.~{Poletto} \& S.~T. {Suess}, 201

\bibitem[{{Gopalswamy} {et~al.}(2003){Gopalswamy}, {Lara}, {Yashiro}, {Nunes},
  \& {Howard}}]{2003iscs.symp..403G}
{Gopalswamy}, N., {Lara}, A., {Yashiro}, S., {Nunes}, S., \& {Howard}, R.~A.
  2003, in ESA SP-535: Solar Variability as an Input to the Earth's
  Environment, ed. A.~{Wilson}, 403--414

\bibitem[{{Howard} {et~al.}(2008){Howard}, {Moses}, {Vourlidas}, {Newmark},
  {Socker}, {Plunkett}, {Korendyke}, {Cook}, {Hurley}, {Davila}, {Thompson},
  {St Cyr}, {Mentzell}, {Mehalick}, {Lemen}, {Wuelser}, {Duncan}, {Tarbell},
  {Wolfson}, {Moore}, {Harrison}, {Waltham}, {Lang}, {Davis}, {Eyles},
  {Mapson-Menard}, {Simnett}, {Halain}, {Defise}, {Mazy}, {Rochus}, {Mercier},
  {Ravet}, {Delmotte}, {Auchere}, {Delaboudiniere}, {Bothmer}, {Deutsch},
  {Wang}, {Rich}, {Cooper}, {Stephens}, {Maahs}, {Baugh}, {McMullin}, \&
  {Carter}}]{2008SSRv..136...67H}
{Howard}, R.~A., {Moses}, J.~D., {Vourlidas}, A., {Newmark}, J.~S., {Socker},
  D.~G., {Plunkett}, S.~P., {Korendyke}, C.~M., {Cook}, J.~W., {Hurley}, A.,
  {Davila}, J.~M., {Thompson}, W.~T., {St Cyr}, O.~C., {Mentzell}, E.,
  {Mehalick}, K., {Lemen}, J.~R., {Wuelser}, J.~P., {Duncan}, D.~W., {Tarbell},
  T.~D., {Wolfson}, C.~J., {Moore}, A., {Harrison}, R.~A., {Waltham}, N.~R.,
  {Lang}, J., {Davis}, C.~J., {Eyles}, C.~J., {Mapson-Menard}, H., {Simnett},
  G.~M., {Halain}, J.~P., {Defise}, J.~M., {Mazy}, E., {Rochus}, P., {Mercier},
  R., {Ravet}, M.~F., {Delmotte}, F., {Auchere}, F., {Delaboudiniere}, J.~P.,
  {Bothmer}, V., {Deutsch}, W., {Wang}, D., {Rich}, N., {Cooper}, S.,
  {Stephens}, V., {Maahs}, G., {Baugh}, R., {McMullin}, D., \& {Carter}, T.
  2008, Space Science Reviews, 136, 67

\bibitem[{{Howard} {et~al.}(1985){Howard}, {Sheeley}, {Michels}, \&
  {Koomen}}]{1985JGR....90.8173H}
{Howard}, R.~A., {Sheeley}, Jr., N.~R., {Michels}, D.~J., \& {Koomen}, M.~J.
  1985, \jgr, 90, 8173

\bibitem[{{Howard} {et~al.}(1986){Howard}, {Sheeley}, {Michels}, \&
  {Koomen}}]{1986shtd.symp..107H}
{Howard}, R.~A., {Sheeley}, Jr., N.~R., {Michels}, D.~J., \& {Koomen}, M.~J.
  1986, in ASSL Vol. 123: The Sun and the Heliosphere in Three Dimensions, ed.
  R.~G. {Marsden}, 107--111

\bibitem[{{Hundhausen}(1993)}]{1993JGR....9813177H}
{Hundhausen}, A.~J. 1993, \jgr, 98, 13177

\bibitem[{{Jackson} \& {Hildner}(1978)}]{1978SoPh...60..155J}
{Jackson}, B.~V. \& {Hildner}, E. 1978, \solphys, 60, 155

\bibitem[{{Jacobs} {et~al.}(2005){Jacobs}, {Poedts}, {Van der Holst}, \&
  {Chan{\'e}}}]{2005A&A...430.1099J}
{Jacobs}, C., {Poedts}, S., {Van der Holst}, B., \& {Chan{\'e}}, E. 2005, \aap,
  430, 1099

\bibitem[{{Kane}(2008)}]{2008SoPh..249..355K}
{Kane}, R.~P. 2008, \solphys, 249, 355

\bibitem[{{Ko} {et~al.}(2003){Ko}, {Raymond}, {Lin}, {Lawrence}, {Li}, \&
  {Fludra}}]{2003ApJ...594.1068K}
{Ko}, Y.-K., {Raymond}, J.~C., {Lin}, J., {Lawrence}, G., {Li}, J., \&
  {Fludra}, A. 2003, \apj, 594, 1068

\bibitem[{{Low}(1982)}]{1982ApJ...254..796L}
{Low}, B.~C. 1982, \apj, 254, 796

\bibitem[{{Low}(1984)}]{1984ApJ...281..392L}
---. 1984, \apj, 281, 392

\bibitem[{{Lu} \& {Hamilton}(1991)}]{1991ApJ...380L..89L}
{Lu}, E.~T. \& {Hamilton}, R.~J. 1991, \apjl, 380, L89

\bibitem[{{MacQueen} {et~al.}(1980){MacQueen}, {Csoeke-Poeckh}, {Hildner},
  {House}, {Reynolds}, {Stanger}, {Tepoel}, \& {Wagner}}]{1980SoPh...65...91M}
{MacQueen}, R.~M., {Csoeke-Poeckh}, A., {Hildner}, E., {House}, L., {Reynolds},
  R., {Stanger}, A., {Tepoel}, H., \& {Wagner}, W. 1980, \solphys, 65, 91

\bibitem[{{Moses} {et~al.}(1997){Moses}, {Clette}, {Delaboudini{\`e}re},
  {Artzner}, {Bougnet}, {Brunaud}, {Carabetian}, {Gabriel}, {Hochedez},
  {Millier}, {Song}, {Au}, {Dere}, {Howard}, {Kreplin}, {Michels}, {Defise},
  {Jamar}, {Rochus}, {Chauvineau}, {Marioge}, {Catura}, {Lemen}, {Shing},
  {Stern}, {Gurman}, {Neupert}, {Newmark}, {Thompson}, {Maucherat},
  {Portier-Fozzani}, {Berghmans}, {Cugnon}, {van Dessel}, \&
  {Gabryl}}]{1997SoPh..175..571M}
{Moses}, D., {Clette}, F., {Delaboudini{\`e}re}, J.-P., {Artzner}, G.~E.,
  {Bougnet}, M., {Brunaud}, J., {Carabetian}, C., {Gabriel}, A.~H., {Hochedez},
  J.~F., {Millier}, F., {Song}, X.~Y., {Au}, B., {Dere}, K.~P., {Howard},
  R.~A., {Kreplin}, R., {Michels}, D.~J., {Defise}, J.~M., {Jamar}, C.,
  {Rochus}, P., {Chauvineau}, J.~P., {Marioge}, J.~P., {Catura}, R.~C.,
  {Lemen}, J.~R., {Shing}, L., {Stern}, R.~A., {Gurman}, J.~B., {Neupert},
  W.~M., {Newmark}, J., {Thompson}, B., {Maucherat}, A., {Portier-Fozzani}, F.,
  {Berghmans}, D., {Cugnon}, P., {van Dessel}, E.~L., \& {Gabryl}, J.~R. 1997,
  \solphys, 175, 571

\bibitem[{{Munro} {et~al.}(1979){Munro}, {Gosling}, {Hildner}, {MacQueen},
  {Poland}, \& {Ross}}]{1979SoPh...61..201M}
{Munro}, R.~H., {Gosling}, J.~T., {Hildner}, E., {MacQueen}, R.~M., {Poland},
  A.~I., \& {Ross}, C.~L. 1979, \solphys, 61, 201

\bibitem[{{{\"O}zg{\"u}{\c c}} \& {Ata{\c c}}(2001)}]{2001IAUS..203..125O}
{{\"O}zg{\"u}{\c c}}, A. \& {Ata{\c c}}, T. 2001, in IAU Symposium, ed.
  P.~{Brekke}, B.~{Fleck}, \& J.~B. {Gurman}, 125

\bibitem[{{Parker}(1983)}]{1983sowi.conf...23P}
{Parker}, E.~N. 1983, in Solar Wind Conference, ed. J.~M. {Wilcox} \& A.~J.
  {Hundhausen}, 23--32

\bibitem[{{Riley} {et~al.}(2007){Riley}, {Lionello}, {Miki{\'c}}, {Linker},
  {Clark}, {Lin}, \& {Ko}}]{2007ApJ...655..591R}
{Riley}, P., {Lionello}, R., {Miki{\'c}}, Z., {Linker}, J., {Clark}, E., {Lin},
  J., \& {Ko}, Y.-K. 2007, \apj, 655, 591

\bibitem[{{Robbrecht} \& {Berghmans}(2004)}]{2004A&A...425.1097R}
{Robbrecht}, E. \& {Berghmans}, D. 2004, \aap, 425, 1097

\bibitem[{{Sheeley} {et~al.}(1997){Sheeley}, {Wang}, {Hawley}, {Brueckner},
  {Dere}, {Howard}, {Koomen}, {Korendyke}, {Michels}, {Paswaters}, {Socker},
  {St.~Cyr}, {Wang}, {Lamy}, {Llebaria}, {Schwenn}, {Simnett}, {Plunkett}, \&
  {Biesecker}}]{1997ApJ...484..472S}
{Sheeley}, Jr., N.~R., {Wang}, Y.-M., {Hawley}, S.~H., {Brueckner}, G.~E.,
  {Dere}, K.~P., {Howard}, R.~A., {Koomen}, M.~J., {Korendyke}, C.~M.,
  {Michels}, D.~J., {Paswaters}, S.~E., {Socker}, D.~G., {St.~Cyr}, O.~C.,
  {Wang}, D., {Lamy}, P.~L., {Llebaria}, A., {Schwenn}, R., {Simnett}, G.~M.,
  {Plunkett}, S., \& {Biesecker}, D.~A. 1997, \apj, 484, 472

\bibitem[{{St.~Cyr} {et~al.}(2000){St.~Cyr}, {Plunkett}, {Michels},
  {Paswaters}, {Koomen}, {Simnett}, {Thompson}, {Gurman}, {Schwenn}, {Webb},
  {Hildner}, \& {Lamy}}]{2000JGR...10518169S}
{St.~Cyr}, O.~C., {Plunkett}, S.~P., {Michels}, D.~J., {Paswaters}, S.~E.,
  {Koomen}, M.~J., {Simnett}, G.~M., {Thompson}, B.~J., {Gurman}, J.~B.,
  {Schwenn}, R., {Webb}, D.~F., {Hildner}, E., \& {Lamy}, P.~L. 2000, \jgr,
  105, 18169

\bibitem[{{Temmer} {et~al.}(2003){Temmer}, {Veronig}, \&
  {Hanslmeier}}]{2003SoPh..215..111T}
{Temmer}, M., {Veronig}, A., \& {Hanslmeier}, A. 2003, \solphys, 215, 111

\bibitem[{{Vanlommel} {et~al.}(2004){Vanlommel}, {Cugnon}, {Linden},
  {Berghmans}, \& {Clette}}]{2004SoPh..224..113V}
{Vanlommel}, P., {Cugnon}, P., {Linden}, R.~A.~M.~V.~D., {Berghmans}, D., \&
  {Clette}, F. 2004, \solphys, 224, 113

\bibitem[{{Vlahos} {et~al.}(2004){Vlahos}, {Isliker}, \&
  {Lepreti}}]{2004ApJ...608..540V}
{Vlahos}, L., {Isliker}, H., \& {Lepreti}, F. 2004, \apj, 608, 540

\bibitem[{{Vourlidas} {et~al.}(2002){Vourlidas}, {Buzasi}, {Howard}, \&
  {Esfandiari}}]{2002svco.conf...91V}
{Vourlidas}, A., {Buzasi}, D., {Howard}, R.~A., \& {Esfandiari}, E. 2002, in
  ESA SP-506: Solar Variability: From Core to Outer Frontiers, ed. A.~{Wilson},
  91--94

\bibitem[{{Vourlidas} {et~al.}(2003){Vourlidas}, {Wu}, {Wang}, {Subramanian},
  \& {Howard}}]{2003ApJ...598.1392V}
{Vourlidas}, A., {Wu}, S.~T., {Wang}, A.~H., {Subramanian}, P., \& {Howard},
  R.~A. 2003, \apj, 598, 1392

\bibitem[{{Wang} \& {Sheeley}(2002)}]{2002ApJ...575..542W}
{Wang}, Y.-M. \& {Sheeley}, Jr., N.~R. 2002, \apj, 575, 542

\bibitem[{{Webb} {et~al.}(2003){Webb}, {Burkepile}, {Forbes}, \&
  {Riley}}]{2003JGRA..108.1440W}
{Webb}, D.~F., {Burkepile}, J., {Forbes}, T.~G., \& {Riley}, P. 2003, Journal
  of Geophysical Research (Space Physics), 108, 1440

\bibitem[{{Webb} \& {Howard}(1994)}]{1994JGR....99.4201W}
{Webb}, D.~F. \& {Howard}, R.~A. 1994, \jgr, 99, 4201

\bibitem[{{Webb} \& {Hundhausen}(1987)}]{1987SoPh..108..383W}
{Webb}, D.~F. \& {Hundhausen}, A.~J. 1987, \solphys, 108, 383

\bibitem[{{Yashiro} {et~al.}(2004){Yashiro}, {Gopalswamy}, {Michalek},
  {St.~Cyr}, {Plunkett}, {Rich}, \& {Howard}}]{2004JGRA..10907105Y}
{Yashiro}, S., {Gopalswamy}, N., {Michalek}, G., {St.~Cyr}, O.~C., {Plunkett},
  S.~P., {Rich}, N.~B., \& {Howard}, R.~A. 2004, Journal of Geophysical
  Research (Space Physics), 109, 7105

\bibitem[{{Yurchyshyn} {et~al.}(2005){Yurchyshyn}, {Yashiro}, {Abramenko},
  {Wang}, \& {Gopalswamy}}]{2005ApJ...619..599Y}
{Yurchyshyn}, V., {Yashiro}, S., {Abramenko}, V., {Wang}, H., \& {Gopalswamy},
  N. 2005, \apj, 619, 599

\bibitem[{{Zhao} {et~al.}(2002){Zhao}, {Plunkett}, \&
  {Liu}}]{2002JGRA.107hSSH13Z}
{Zhao}, X.~P., {Plunkett}, S.~P., \& {Liu}, W. 2002, Journal of Geophysical
  Research (Space Physics), 107, 13

\end{thebibliography}

\end{document}